\documentclass[%
 reprint,
superscriptaddress,
 amsmath,amssymb,
prd,
]{revtex4-1}

\usepackage[dvipdfmx]{graphicx}% Include figure files
\usepackage{dcolumn}% Align table columns on decimal point
\usepackage{bm}% bold math
\usepackage{float}
\usepackage{color}

\newcommand{\msun}{\mathrm{M_\odot}}
\newcommand{\zsun}{\mathrm{Z_\odot}}
\newcommand{\tsun}{\mathrm{T_\odot}}
\newcommand{\mcz}{\mathcal{M}}

\begin{document}

%\preprint{APS/123-QED}

\title{How to confirm the existence of population III stars \\ by observations of gravitational waves}

\author{Akinobu Miyamoto}
\email{amiyamoto@yukimura.hep.osaka-cu.ac.jp}
 \affiliation{Department of Physics, Graduate School of Science, Osaka city University, Osaka 558-8585, Japan}
\author{Tomoya Kinugawa}
 \affiliation{Institute for Cosmic Ray Research, University of Tokyo, Chiba 277-8582, Japan}
\author{Takashi Nakamura}
 \affiliation{Department of Physics, Graduate School of Science, Kyoto University, Kyoto 606-8502, Japan}
\author{Nobuyuki Kanda}
\affiliation{Department of Physics, Graduate School of Science, Osaka city University, Osaka 558-8585, Japan}

%%%%%%%%  Abstract  %%%%%%%%
\begin{abstract}
We propose a method for confirmation of the existence of Population~III (Pop~III) stars with massive black hole binaries as GW150914 in gravitational wave (GW) observation. 
When we get enough number of events, we want to determine which model is closer to reality, with and without Pop~III stars.
We need to prepare various ``Pop~I/II models'' and various ``Pop~I/II/III models'' and investigate which model is consistent with the events.
To demonstrate our analysis, we simulate detections of GW events for some examples of population synthesis models with and without Pop~III stars.
We calculate the likelihood ratio with the realistic number of events and evaluate the probability of identifying the existence of Pop~III stars.
In typical cases, our analysis can distinguish between Pop~I/II model and Pop~I/II/III model 
with 90\% probability by 22 GW signals from black hole-black hole binary mergers.\\


%\noindent Revised at \today
\date{\today}% It is always \today, today,

\end{abstract}
\maketitle

%\tableofcontents

%%%%%%%%  Introduction  %%%%%%%%
\section{Introduction}
%%  [LIGO O1 run, GW150914, Pop III] %%
In 2015, Advanced LIGO \cite{LIGOpaper} detected gravitational waves (GWs) in its first observational run (O1 run). 
The first detected event GW150914 is a black hole-black hole (BH-BH) binary whose masses are $36~\msun$ and $29~\msun$ \cite{abbott2016}.
To explain the existence of such high mass black holes, there are some scenarios. One of the possible origins of GW150914 is the Population~III (Pop~III) stars, i.e., the zero metal stars formed first after the big bang \cite{kinugawa2014,kinugawa2016}. 
While Population~I (Pop~I) stars and Population~II (Pop~II) stars with nonzero metal are observed electromagnetically, the existence of Pop~III stars has not been confirmed yet by electromagnetic wave observations. 

%%  [binary parameter, LIGO, Virgo, KAGRA] %%
However, the GW observation with enough number of events may confirm the existence of Pop~III star.
From the detection of GW from a compact star binary coalescence, we can obtain binary parameters such as masses, distance, redshift, sky location, inclination of orbit, spins and eccentricity of the compact binary.
Second generation GW detectors such as Advanced LIGO, Advanced Virgo \cite{Virgopaper} and KAGRA \cite{kagra} would detect many high mass BH-BH binary mergers and provide us with distributions of the binary parameters.
The GW from such high mass BH-BH binary mergers will give us information related to binary evolution, binary formation scenario and Pop~III star.

%%  [likelihood ratio, model  {discrimination}] %%
When we observe enough number of events, we want to determine which model is closer to reality and to know whether the plausible model consists of Pop~III stars or not.
To do that, we need to prepare various Pop~I/II (combination of Pop~I and Pop~II)  models and various Pop~I/II/III (combination of Pop~I/II and Pop~III) models and investigate which model is consistent with the events.
In order to investigate whether the plausible model consists of Pop~III stars or not, we adopt the likelihood analysis of mass distributions of population synthesis simulations. Then we evaluate the probability of identifying the existence of Pop~III stars.
In this paper, we demonstrate our analysis method with examples of population synthesis models.

%In our study, we ignored  contribution of the 
There are another binary formation scenarios such as primordial BH formation \cite{nakamura1997, sasaki2016} and dynamical binary formation in a globular cluster \cite{portegies2000}.
To demonstrate our analysis, however, we only calculate Pop~III and Pop~I/II cases as an example.
In future we will take the primordial BH formation and the dynamical binary formation in a globular cluster into account \cite{zevin2017}.

%%  [organization of paper] %%
This paper is organized as follows.
In Sec.~\ref{sec:method}, we explain the examples of population synthesis models, 
a method of Monte Carlo simulations of GW detections from BH-BH binaries
and a method for confirmation of the existence of Pop~III stars.
In Sec.~\ref{sec:results},  we show results of simulations of the GW detections and 
a probability of distinguishing between Pop~I/II model and Pop~I/II/III model.
Discussion and summary are shown in Sec.~\ref{sec:summary}.

%%%%%%%%  analysis method  %%%%%%%%
\section{analysis method}
\label{sec:method}

%%%%%%%%  Example population synthesis models  %%%%%%%%
\subsection{Example of population synthesis models}
\label{sec:ex-models}

%%  [Example models of population synthesis, Pop I/II, Pop III] %%

In order to simulate the GW events from the BH-BH binary mergers, we use the results of the binary population synthesis.
The binary population synthesis is the Monte Carlo simulation.
In this simulation, we prepare the initial parameters such as eccentricity, semimajor axis, total mass, and the mass ratio of the Pop~III binary following the given probability distribution functions.  The population synthesis simulation can estimate the merger rate density as a function of redshift and the binary mass spectrum, which strongly depend on metallicity of stars.
In Refs.~\cite{kinugawa2014, kinugawa2016}, Pop~III binaries tend to become BH-BH binaries which
merge within the Hubble time.
The typical mass of Pop~III BH-BH binaries is $\sim30~\msun$ which does not depend on the binary parameters and the initial distribution functions.
On the other hand, in the case of Pop~I and Pop~II stars, the typical mass of BHs is $10~\msun$ or so although there is the model dependence \cite{dominik2012, dominik2013}.
We mention that the purpose of this paper is propose the method for confirmation of the existence of Pop~III stars, not to verify a specific population synthesis model.

In Refs.~\cite{dominik2012} and \cite{dominik2013}, the authors discussed 
four models of Ref.~\cite{dominik2012}. In this paper, we employ the four models in Ref.~\cite{dominik2012} as examples of models.
Here we use one of the four examples of models for our analysis. Then the remains of the models are used in Sec.~\ref{sec:summary}.
As an example of Pop~I/II model,
we employ Dominik's standard model submodel B (standard) with metallicity of $Z=\zsun$ and $Z=0.1~\zsun$ \cite{dominik2012}.
They use initial mass function (IMF) $\Psi(m) \propto m^{-2.7}~(1.0~\msun \leq m < 150~\msun)$ for Pop~I/II model, where $m$ is the mass of the zero age main sequence star.
In their simulation, they calculated galactic merger rates of the local universe. To calculate a merger rate density within a few Gpc cubic, we employ the density of galaxies $\rho_\mathrm{gal} = 0.0116 ~\mathrm{Mpc}^{-3}$ \cite{kopparapu2008}.
We restrict Dominik's Pop~I/II BH-BH binaries that merge within 10 Gyrs.
Their population synthesis data are available on their online database \url{http://www.syntheticuniverse.org}.

As an example of Pop~III model, we use Kinugawa's models \cite{kinugawa2016}.
An employed model in this paper is the standard model (the flat initial mass function model) with the heavier initial mass range and the conservative core-merger criterion during the common envelope phase. The initial mass range of the Pop~III model is $10~\msun < m < 140~\msun$, where $m$ is the mass of the zero age main sequence star.
The core-merger criterion of the common envelope phase in the Pop~III model is a conservative one which is the same as that of Belczynski and Dominik \cite{belczynski2002, kinugawa2014}. We call the standard model as ``IMF:Flat'' in this paper.

In Fig.~\ref{f:nd-zmc-1model}, dotted (blue) and solid (red) lines correspond to average distributions of redshifted chirp mass of simulated BH-BH binary detections in the Pop~I/II model and the Pop~III model, respectively.
The redshifted chirp mass distributions of Pop~I/II model and Pop~III model have peaks at $\sim 10~\msun$ and $\sim 30~\msun$, respectively.

%%%%%%%%  Simulation of detection of GW  %%%%%%%%
\subsection{Simulation of GW detections}
\label{sec:simulation}
%%  [binary parameters, cumulative merger rate, lambda CDM model] %%

We generate binary parameters such as redshift $z$, each mass $m_1, m_2$, right ascension $\alpha$, declination $\delta$ of the binary, inclination angle $\iota$ of orbit and polarization angle $\psi$ of GW.
We use the population synthesis models for distributions of $z, m_1$, and $m_2$.
We perform the Monte Carlo simulations to generate $\alpha, \delta, \iota$ and $\psi$ using the isotropic distributions.
The cumulative merger rate as a function of redshift is given by
\begin{equation}
N(z) = 4\pi \int_0^z R(z') r^2(z') \frac{1}{1+z'} \frac{dr(z')}{dz'} dz',
\end{equation}
where $R(z)$ is the merger rate density per comoving volume calculated from population synthesis models, $1/(1+z)$ is the cosmological time dilation effect and $r(z)$ is the comoving distance, which  is given by
\begin{equation}
r(z) = \frac{c}{H_0} \int_0^z \frac{dz'}{\sqrt{\Omega_\mathrm{m}(1+z')^3 + \Omega_\Lambda}},
\end{equation}
where we adopt the $\Lambda$-CDM cosmological model \cite{spergel2007} and $c, H_0, \Omega_\mathrm{m}$, and $\Omega_\Lambda$ are the speed of light, the present Hubble parameter, the matter density parameter, and the dark energy parameter, respectively. We adopt Planck collaboration 2013 values for the cosmological parameters \cite{planck2013}.

%%  [signal to noise ratio, inspiral, KAGRA sensitivity, S/Nth=8] %%
In this paper, for simplicity, we consider only the inspiral phase of GWs.
Since the gravitational waveform of coalescing compact binaries can be predicted \cite{blanchet1996}, we calculate a signal-to-noise ratio (S/N) of the inspiral GW by following equations \cite{sathyaprakash2009},
\begin{align}
(\mathrm{S/N})^2
=& 4  \int_{f_\mathrm{min}}^{f_\mathrm{max}} df
\frac{|\tilde{h}(f)|^2}{S_n(f)}  \nonumber \\
=& \frac{5}{6} \frac{\pi^{-4/3} c^2 \ \tsun^{5/3} }{d_\mathrm{L}^2} 
  \left( \frac{\mcz}{\msun} \right)^{5/3} I ~ \beta,    \label{e:snr}  \\
I =& \int_{f_\mathrm{min}}^{f_\mathrm{max}} df
\frac{f^{-7/3}}{S_n(f)} ,  \\
\beta =& \left(\frac{1+\cos^2 \iota}{2}\right)^2 ~F_+^2 + \cos^2 \iota ~F_\times^2, \label{e:beta}\\
\mcz=&(1+z)(m_1+m_2)\left[ \frac{m_1 m_2}{(m_1+m_2)^2}\right]^{3/5}, 
\label{e:mcz}
\end{align}
where $\tilde{h}(f), S_n(f), f_\mathrm{min}$, $f_\mathrm{max}$ and $\mcz$ denote the Fourier transform of the inspiral GW signal $h(t)$, the noise power spectral density of KAGRA \cite{kagra}, the low frequency cutoff, the high frequency cutoff, and the redshifted chirp mass, respectively. The $f_\mathrm{max}$ is the GW frequency at the ISCO (innermost stable circular orbit) given by $f_\mathrm{max}=f_\mathrm{ISCO} = 1/[6^{3/2}\pi \tsun (1+z)(m_1+m_2)/\msun]$, where $\tsun = G\msun/c^3$.
We use $f_\mathrm{min}=1\rm ~Hz$ from the noise curve data of KAGRA \cite{kagra}.
The luminosity distance $d_\mathrm{L}$ is given by $(1+z) r(z)$.
$F_+$ and $F_\times$ are detector antenna patterns.
The detection criterion is settled as S/N $\ge$ 8.

To validate Eq.~(\ref{e:snr})-(\ref{e:beta}) which calculate S/N of a inspiral GW,  
we calculate S/N of the inspiral phase of a GW150914-like BH-BH binary
which is parametrized as : $m_1=36~\msun, m_2=29~\msun$, $z=0.09$ \cite{abbott2016}, inclination angle of orbit $\iota=150^\circ$ from Fig.~2 of Ref.~\cite{propertygw150914} and sky location of the binary at the time of detection are $\alpha=-30^\circ$ and $\delta=-75^\circ$ from the right plot of Fig.~6 of Ref.~\cite{O1run}.
$\psi$ is assumed to be 0.
The detector sensitivity is assumed to be the sensitivity of LIGO O1 run \cite{abbott2016}.
The S/N calculated by Eq.~(\ref{e:snr})-(\ref{e:beta}) with Hanford and Livingston sensitivities are 16.4 and 9.8, respectively.
To check the consistency, we calculated S/N of quasinormal mode (QNM) using Eq.~(B14) of Ref.~\cite{Flanagan1998}, then we combined the S/N of inspiral and QNM. 
The final BH Kerr parameter and fraction of radiated energy in ringdown phase in Eq.~(B14) of Ref.~\cite{Flanagan1998} are assumed to be 0.69 and 0.03, respectively.
As a result, S/N of QNM GW of Hanford and Livingston are 9.5 and 8.0, respectively.
Then the quadrature sum of the signal-to-noise ratios of inspiral and ringdown in Hanford and Livingston are 19.0 and 12.7, respectively.
Combined signal-to-noise ratio of two detectors is $\sqrt{19.0^2+12.7^2} = 22.8$.
While LIGO observed GW150914 with a combined signal-to-noise ratio of 24.
The S/N calculated by Eq.~(\ref{e:snr})-(\ref{e:beta}) and Eq.~(B14) of Ref.~\cite{Flanagan1998} is similar to the LIGO result.

%%  [  Figure 1 ]  %%
\begin{figure}
	\includegraphics[width=8.5cm]{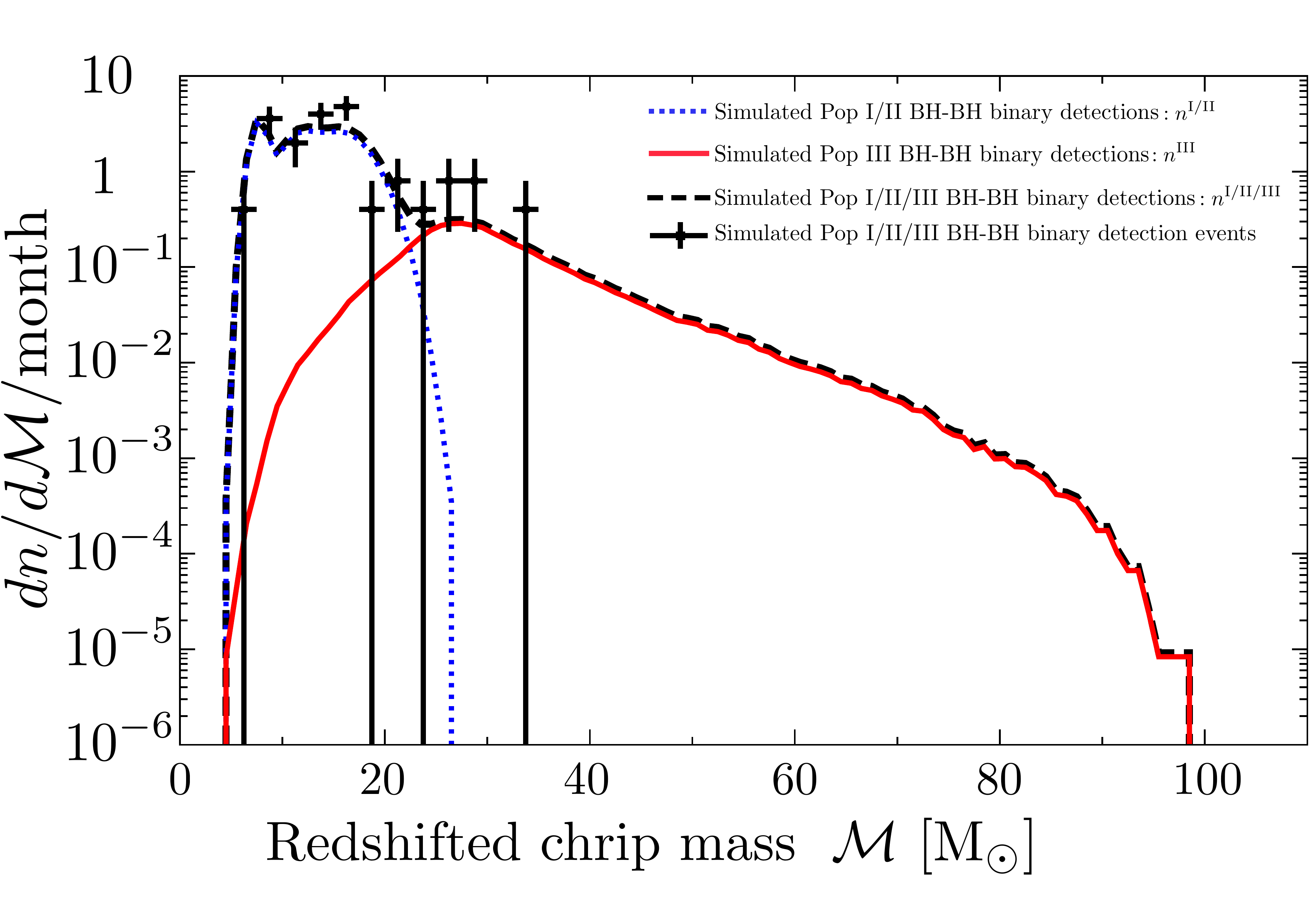}
	\caption{
Average distributions of redshifted chirp mass $\mcz$ of simulated BH-BH binary detections for a 1-month period, and a simulated observation result for a 1-month period of GW detection simulation in KAGRA.
The horizontal axis is the redshifted chirp mass. 
The vertical axis is the detection rate per redshifted chirp mass of BH-BH binary mergers [/$\msun$/month].
The dotted (blue) and solid (red) lines correspond to the average distributions of redshifted chirp mass $\mcz$ of simulated BH-BH binary detections for a 1-month period of Pop~I/II Standard model and that of Pop~III IMF:Flat, respectively.
The dashed (black) line corresponds to sum of the Pop~I/II model and the Pop~III model, i.e., Pop~I/II/III model.
The crosses and its error bars show the simulated BH-BH binary detection events and its statistical errors on a 1-month period simulation, respectively.
}
	\label{f:nd-zmc-1model}
\end{figure}

In Fig.~\ref{f:nd-zmc-1model}, we show average distributions of redshifted chirp mass $\mcz$ of simulated BH-BH binary detections for a 1-month period, and a simulated observation result for a 1-month period of GW detection simulation in KAGRA.
The horizontal axis is redshifted chirp mass.
The vertical axis is the detection rate per redshifted chirp mass of BH-BH binary mergers [/$\msun$/month].
We define $n^\mathrm{I/II}$, $n^\mathrm{III}$, and $n^\mathrm{I/II/III}$ as the numbers of detections of simulated Pop~I/II BH-BH binary mergers, simulated Pop~III BH-BH binary mergers and simulated Pop~I/II/III BH-BH binary mergers, respectively.
The dotted (blue) and solid (red) lines of Fig.~\ref{f:nd-zmc-1model} are $d n^\mathrm{I/II}/d \mcz$/month which is the average distribution of redshifted chirp mass $\mcz$ of simulated BH-BH binary detections for a 1-month period of Pop~I/II Standard model and $d n^\mathrm{III}/d \mcz$/month which is that of Pop~III IMF:Flat, respectively.
The dashed (black) line in Fig.~\ref{f:nd-zmc-1model} corresponds to $d n^\mathrm{I/II/III}/d \mcz$/month which is the sum of the Pop~I/II model and the Pop~III model, i.e., the Pop~I/II/III model.
The crosses and its error bars in Fig.~\ref{f:nd-zmc-1model} show the simulated BH-BH binary detection events and its statistical errors on a 1-month period simulation, respectively.

From Fig.~\ref{f:nd-zmc-1model}, it is not trivial whether we can distinguish between the redshifted chirp mass distribution of Pop~I/II model and that of Pop~I/II/III model.
We demonstrate a likelihood analysis using simulated observation results and evaluate the probability of identifying the existence of Pop~III stars.
To generate a lot of simulated observation results, we use probability density functions calculated from $d n^\mathrm{I/II}/d \mcz$/month and $d n^\mathrm{I/II/III}/d \mcz$/month in Fig.~\ref{f:nd-zmc-1model}.
 Then we generate realized events following the probability density functions.
Using this generation method of realized events, we can generate a data set with the arbitrary number of simulated BH-BH binary detections.

%%%%%%%%  Likelihood analysis method  %%%%%%%%
\subsection{Likelihood analysis method}
\label{sec:likelihood}
%%  [mass of Pop III BBH tend to be massive than that of Pop I/II BBH, 2 situations, redshifted chirp mass]  %%
According to the results of population synthesis simulations, the Pop~III origin BH-BH binaries tend to be heavier than the Pop~I/II origin ones.
We evaluate the difference in the redshifted chirp mass distribution between a Pop~I/II model and a Pop~I/II/III model.
Pop~I/II BH-BH binaries must exist, so there are two situations:
(1) There are no Pop III BH-BH binary signals and
(2) There are Pop~III BH-BH binary signals.
In the following, since the estimation accuracy of redshifted chirp mass is usually better than that of the other parameters \cite{creightontext}, we employ redshifted chirp mass $\mcz$ in our analysis.

%%  [definition of likelihood, definition of likelihood ratio]  %%
First, we define a data set of redshifted chirp masses of simulated BH-BH binary detections as 
\begin{align}
\vec{\bf{\mcz}}(n) = \{ \mcz_1, \mcz_2, \dotsc, \mcz_n \} ,
\end{align}
where each $\mcz_i ~ (i=1,2,\dotsc,n)$ is a redshifted chirp mass of simulated BH-BH binary merger, respectively. $n$ is the number of simulated BH-BH binary detections.
From the data set, we calculate likelihoods, defined by 
\begin{align}
L(\vec{\bf{\mcz}}(n)|\theta^\mathrm{I/II}) &= \prod_{i=1}^{n} p^\mathrm{I/II}(\mcz_i),\\
L(\vec{\bf{\mcz}}(n)|\theta^\mathrm{I/II/III}) &= \prod_{i=1}^{n} p^\mathrm{I/II/III}(\mcz_i),
\end{align}
where $\theta^\mathrm{I/II}$ and $\theta^\mathrm{I/II/III}$  are parameters of displaying the situations
(1) There are no Pop~III BH-BH binary signals and
(2) There are Pop~III BH-BH binary signals, respectively.
$p^\mathrm{I/II}(\mcz)$ and $p^\mathrm{I/II/III}(\mcz)$ are probability density functions that are defined by 
\begin{align}
p^\mathrm{I/II}(\mcz) &=
 	 \frac{ \frac{d n^\mathrm{I/II}}{d\mcz}(\mcz) }
		 {\int_0^\infty  \frac{d n^\mathrm{I/II}}{d\mcz} (\mcz)  ~d\mcz}, 
\label{e:pdf1}\\
p^\mathrm{I/II/III}(\mcz) &=
	 \frac{ \frac{d n^\mathrm{I/II/III}}{d\mcz}(\mcz) }
		 { \int_0^\infty \frac{d n^\mathrm{I/II/III}}{d\mcz} (\mcz) ~d\mcz},
\label{e:pdf2}
\end{align}
where $n^\mathrm{I/II}$ and $n^\mathrm{I/II/III}$ are the numbers of detections corresponding to dotted (blue) line and dashed (black) line of Fig.~\ref{f:nd-zmc-1model}, respectively.

We define $\mcz_\mathrm{crit}^\mathrm{I/II}$ as the critical cutoff $\mcz$ of detectable Pop~I/II BH-BH binary mergers.
$\mcz_\mathrm{crit}^\mathrm{I/II}$ is determined by original data which is population synthesis data does not a mass distribution, but component masses $m_1$ and $m_2$.
There is maximum chirp mass when we use the data. We defined the $\mcz_\mathrm{crit}^\mathrm{I/II}$ as the redshifted chirp mass where the distribution is sharp drop such as in Fig.~\ref{f:nd-zmc-1model}.
In the case of the standard model and IMF:Flat, $\mcz_\mathrm{crit}^\mathrm{I/II}$ is $\sim27~\msun$ from Fig.~\ref{f:nd-zmc-1model}.
However $\mcz_\mathrm{crit}^\mathrm{I/II}$ depends on the Pop~I/II model.
If a data set has a BH-BH binary whose redshifted chirp mass is more massive than $\mcz_\mathrm{crit}^\mathrm{I/II}$,
a Pop~I/II model and a Pop~I/II/III model are clearly distinguished without performing our likelihood analysis.
If a data set does not satisfy above situation, we have to perform our likelihood analysis.
In the case of the standard model and IMF:Flat, if there are the enough number of BH-BH binary detections to detect 2 or 3 Pop~III BH-BH binary mergers,
we can detect about one Pop~III BH-BH binary whose redshifted chirp mass is larger than $27~\msun$.
Then identifying the existence of Pop~III stars can be easy.
However the larger $\mcz_\mathrm{crit}^\mathrm{I/II}$, the more difficult it is that 
a Pop~I/II model and a Pop~I/II/III model are clearly distinguished.
In the case of Pop~I/II model with large $\mcz_\mathrm{crit}^\mathrm{I/II}$, our likelihood analysis is necessary.
We calculate a log-likelihood ratio
\begin{align}
\ln \Lambda(\vec{\bf{\mcz}}(n)) = \ln \left[ \frac{L(\vec{\bf{\mcz}}(n)|\theta^\mathrm{I/II/III})}{L(\vec{\bf{\mcz}}(n)|\theta^\mathrm{I/II})}\right].
\label{e:likelihood}
\end{align}

%%  [likelihood ratio threshold, p.d.f. of parameter]  %%
Even if there is no Pop III BH-BH binary detection in an observation, the observation data possibly looks like there are Pop III BH-BH binary detections and the log-likelihood ratio $\ln \Lambda(\vec{\bf{\mcz}}(n))$ becomes high.
To avoid such misunderstanding, we have to introduce a false probability which is misidentifying the Pop~III BH-BH binary detection.
We determine a log-likelihood ratio threshold $\ln \Lambda_\mathrm{th}$ as defined below by 1\% false probability.
In order to calculate the log-likelihood ratio threshold $\ln \Lambda_\mathrm{th}$, we consider first situation
(1) There are no Pop~III BH-BH binary signals.
We generate $10^7$ sets of $\vec{\bf{\mcz}}^\mathrm{I/II}(n)$, where $\vec{\bf{\mcz}}^\mathrm{I/II}(n)$ denotes a data set of redshifted chirp masses of simulated Pop~I/II BH-BH binary detections. 
We make a log-likelihood ratio $\ln \Lambda(\vec{\bf{\mcz}}^\mathrm{I/II}(n))$ distribution. The log-likelihood ratio threshold $\ln \Lambda_\mathrm{th}$ is set by 1\% false probability,
\begin{align}
\int_{-\infty}^{\ln \Lambda_\mathrm{th}} 
      p\left[\ln \Lambda(\vec{\bf{\mcz}}^\mathrm{I/II}(n))\right] d(\ln \Lambda)=0.99,
\label{e:threshold-define}
\end{align}
where $p[\ln \Lambda(\vec{\bf{\mcz}}^\mathrm{I/II}(n))]$ is the probability density function of the log-likelihood ratio $\ln \Lambda(\vec{\bf{\mcz}}^\mathrm{I/II}(n))$.
If a data set satisfies the following condition
\begin{align}
 \ln \Lambda(\vec{\bf{\mcz}}(n)) > \ln \Lambda_\mathrm{th},
\label{e:condition2}
\end{align}
the Pop~I/II model and the Pop~I/II/III model are distinguished under the two situations with 1\% false probability.

%% [definition of probability P] %%
To evaluate the probability of identifying the existence of Pop~III stars, we assume the second situation (2) There are Pop~III BH-BH binary signals.
In order to study the dependence of the probability of identifying the existence of Pop III stars on the number of simulated BH-BH binary detections, we simulate the cases that $n$ is from 1 to 250. The maximum number of simulated BH-BH binary detections $n=250$ is determined by the value that a probability of identifying the existence of Pop~III stars as defined below in Eq.~(\ref{e:probability})
 is saturated.
For each $n$, we generate $D=10^7$ sets of $\vec{\bf{\mcz}}^\mathrm{I/II/III}(n)$, where $\vec{\bf{\mcz}}^\mathrm{I/II/III}(n)$ denotes a data set of redshifted chirp masses of simulated Pop~I/II/III BH-BH binary detections. 
We define the number of data sets that Pop~I/II model and Pop~I/II/III model are clearly distinguished as $D_\mathrm{C}$. We define the number of data sets that Pop~I/II model and Pop~I/II/III model are distinguished by the likelihood analysis as $D_\mathrm{L}$.
The probability of identifying the existence of Pop~III stars $P$ is defined by 
\begin{align}
P = \frac{ D_\mathrm{C} + D_\mathrm{L} }{ D }.
\label{e:probability}
\end{align}

%%%%%%%%  Results  %%%%%%%%
\section{Results}
\label{sec:results}

%%%%%%%%  Detection rate and mass distribution  %%%%%%%%
\subsection{Detection rates and redshifted chirp mass distributions}

%%  [simulation of detection of GWs, Kinugawa et al., detection rate, Table 1] %%
We estimated detection rates and average redshifted chirp mass distributions, and generated a simulated observation result using the simulation of GW detections.
We performed the simulation of detection of inspiral GWs for $10^4$ years period of observation.
Detection rates of the standard model and IMF:Flat in KAGRA are calculated as 31.99 /month and 4.685 /month, respectively.
The other second generation GW detectors such as Advanced LIGO and Advanced Virgo can detect as many as the number of BH-BH binaries which are detected by KAGRA.
Detection rates of Pop~III models are shown in Table~18-23 of Ref.~\cite{kinugawa2016}.

%%  [mean distribution of (1+z)Mc, example data by simulation, p.d.f of parameter, Figure 1 ] %%

In Fig.~\ref{f:nd-zmc-1model}, we show average distributions of redshifted chirp mass $\mcz$ of 
simulated BH-BH binary detections for a 1-month period, and a simulated observation result for a 1-month period of GW detection in KAGRA. 
See Sec.~\ref{sec:simulation} for a more detailed explanation.

%\clearpage

%%%%%%%%  Evaluating probability of  {discrimination} between Pop~I/II model and Pop~III model  %%%%%%%%
%\subsection{Evaluating probability of  {discrimination} between Pop~I/II model and Pop~I/II/III model}
\subsection{Evaluating probability of identifying the existence of Pop~III stars}
%%  [probability of  {discrimination} for 1 month simulation, Figure 2, Table II ] %%

%%  [  Figure 2 ]  %%
\begin{figure}
	\includegraphics[width=8.5cm]{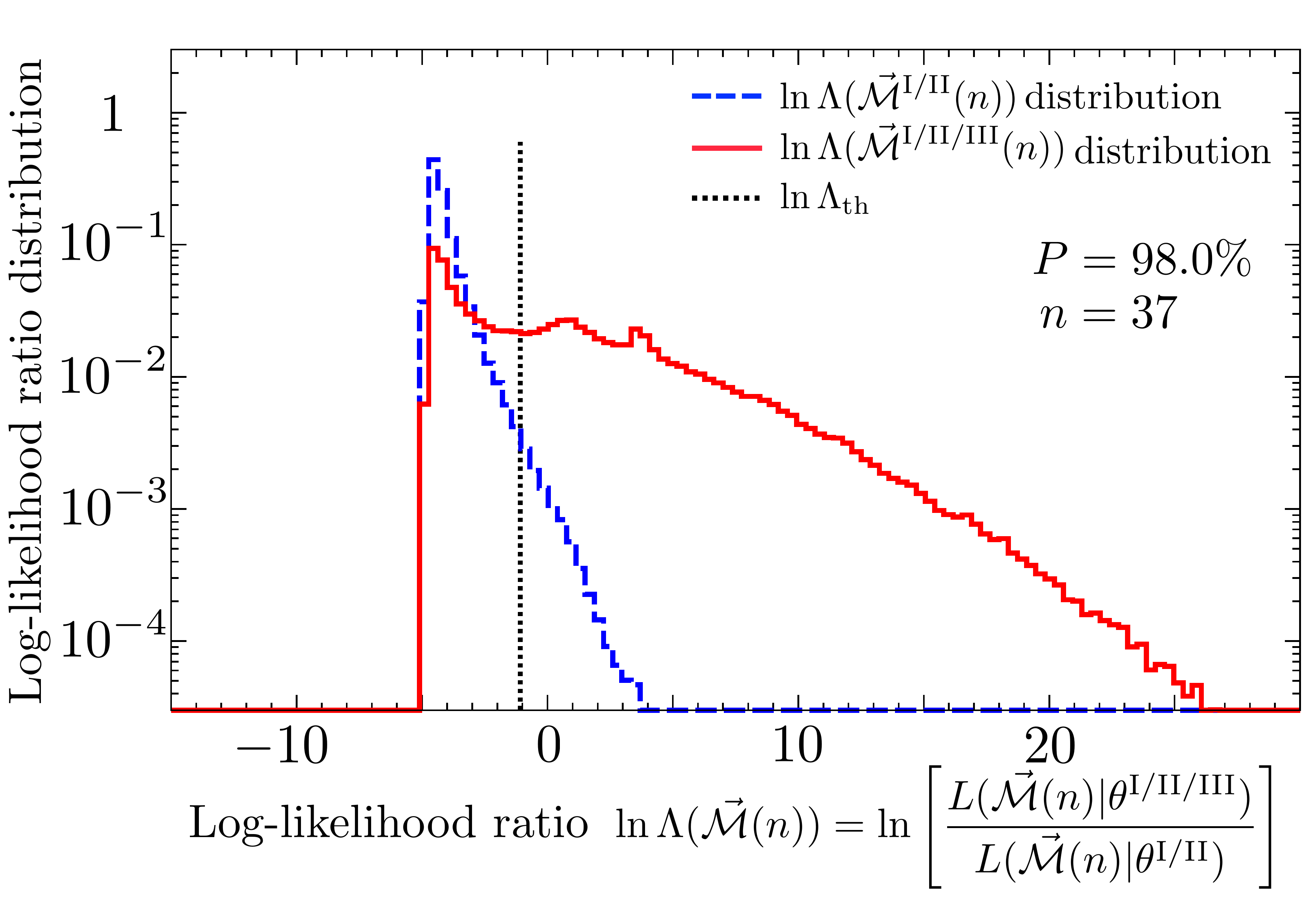}
	\caption{
Distributions of the log-likelihood ratio $\ln \Lambda(\vec{\bf{\mcz}})$ for equivalent a 1-month period of observation.
We use the detection rate of Pop~I/II/III BH-BH binary mergers rounded off to the nearest integer as the value of $n$. In the case of the combination of standard model and IMF:Flat, the detection rate of $31.99+4.685=36.675$ is rounded off to the nearest integer as $n=37$.
The horizontal axis is the log-likelihood ratio. The vertical axis is the probability distribution function of the log-likelihood ratio.
The dashed (blue) and solid (red) lines represent distributions of the $\ln \Lambda(\vec{\bf{\mcz}}^\mathrm{I/II})$ and 
 $\ln \Lambda(\vec{\bf{\mcz}}^\mathrm{I/II/III})$, respectively.
The dotted (black) line shows the log-likelihood ratio threshold $\ln \Lambda_\mathrm{th}$.
The probability of identifying the existence of Pop~III stars $P$ defined in Eq.~(\ref{e:probability}) is estimated to be 98.0\% at $n=37$ events.
}
	\label{f:likelihood-hist-1mon-1model}
\end{figure}

We evaluated the probability of identifying the existence of Pop~III stars.
Figure~\ref{f:likelihood-hist-1mon-1model} shows a result of the likelihood analysis for equivalent a 1-month period of observation.
We use the detection rate of Pop~I/II/III BH-BH binary mergers rounded off to the nearest integer as the value of $n$. In the case of the combination of standard model and IMF:Flat, the detection rate of $31.99+4.685=36.675$ is rounded off to the nearest integer as $n=37$.

In Fig.~\ref{f:likelihood-hist-1mon-1model}, we show the distribution of the log-likelihood ratio $\ln \Lambda(\vec{\bf{\mcz}}^\mathrm{I/II}(n))$ (dashed blue), the log-likelihood ratio threshold $\ln \Lambda_\mathrm{th}$ (dotted black) and the distribution of the log-likelihood ratio $\ln \Lambda(\vec{\bf{\mcz}}^\mathrm{I/II/III}(n))$ (solid red)  for a combination of models of Pop~I/II standard model and Pop~III IMF:Flat.
In the case of (1) there are no Pop III BH-BH binary signals, the log-likelihood ratio distribution becomes 
the dashed (blue) line in Fig.~\ref{f:likelihood-hist-1mon-1model}.
To distinguish between Pop~I/II model and Pop~I/II/III model, we set the log-likelihood ratio threshold (dotted black) with 1\% false probability.
In the case of the second situation (2) there are Pop~III BH-BH binary signals, the log-likelihood ratio distribution becomes the solid (red) line in Fig.~\ref{f:likelihood-hist-1mon-1model}.
As a result, the number $D_C =$ 9 503 988 and the number $D_L =$ 294 964.
The probability of identifying the existence of Pop~III stars $P$ is estimated to be 98.0\% at $n=37$ events.

%%  [  Figure 3 ]  %%
\begin{figure}
\vspace{0.1cm}
	\includegraphics[width=8.5cm]{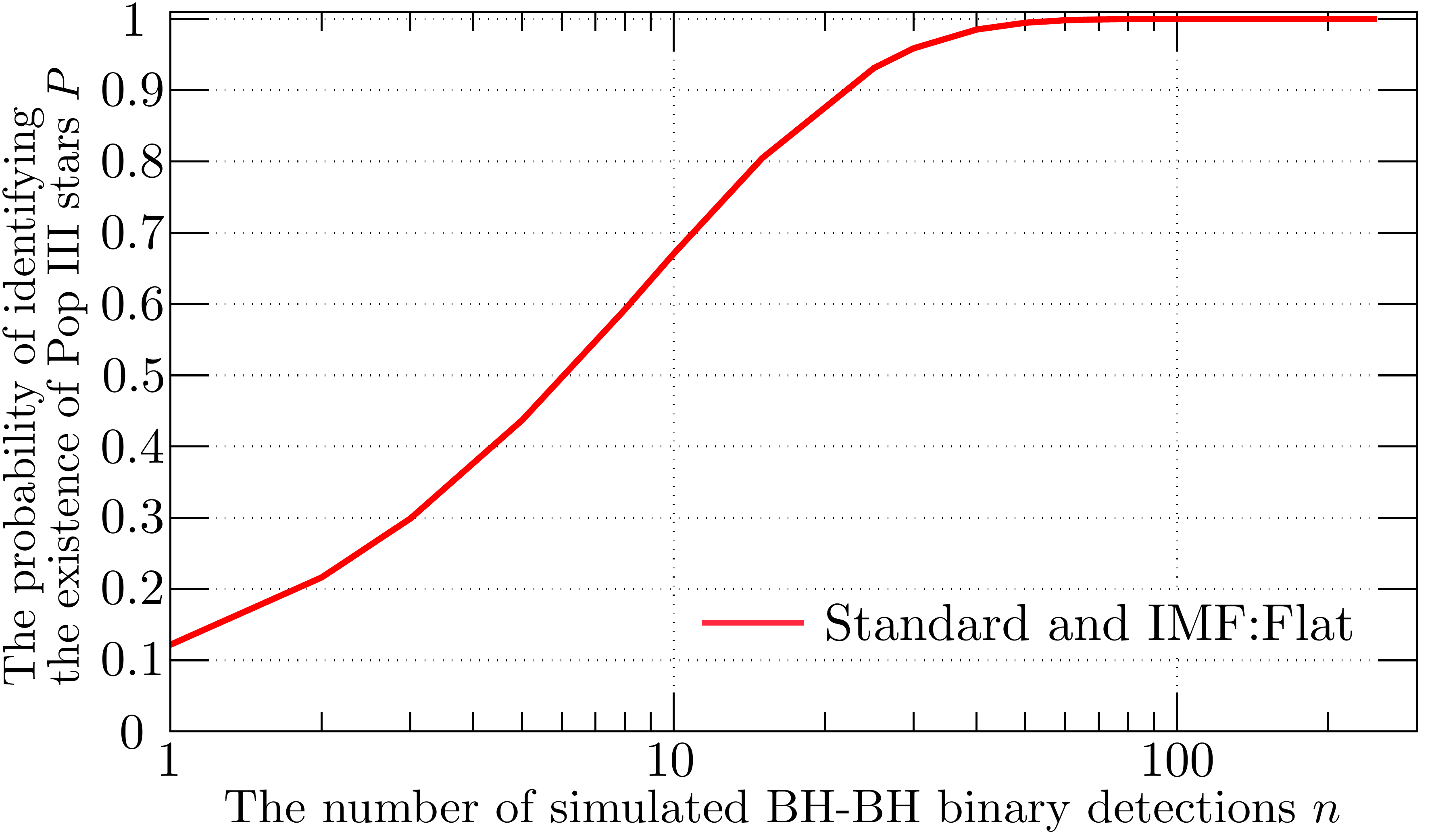}
\caption{
The probability of identifying the existence of Pop~III stars $P$ as a function of the number of simulated BH-BH binary detections $n$.
The horizontal axis is the number of simulated BH-BH binary detections $n$. The vertical axis shows the probability of identifying the existence of Pop~III stars $P$. 
}
\label{f:num-prob-P-1model}
\end{figure}

%%  [ probability of  {discrimination} as a function of #, Figure 3, model dependence, conclusion] %%
Next we show the dependence of the probability of identifying the existence of Pop~III stars $P$ on the number of simulated BH-BH binary detections $n$.
Figure~\ref{f:num-prob-P-1model} shows the $P$ as a function of the number of simulated BH-BH binary detections $n$.
The $P$ is growing with increasing the number of simulated BH-BH binary detections $n$.
The $P$ reaches 90\% at $n=22$ events.

%\clearpage

%%  [  Figure 4 ]  %%
\begin{figure*}
	\includegraphics[width=15cm]{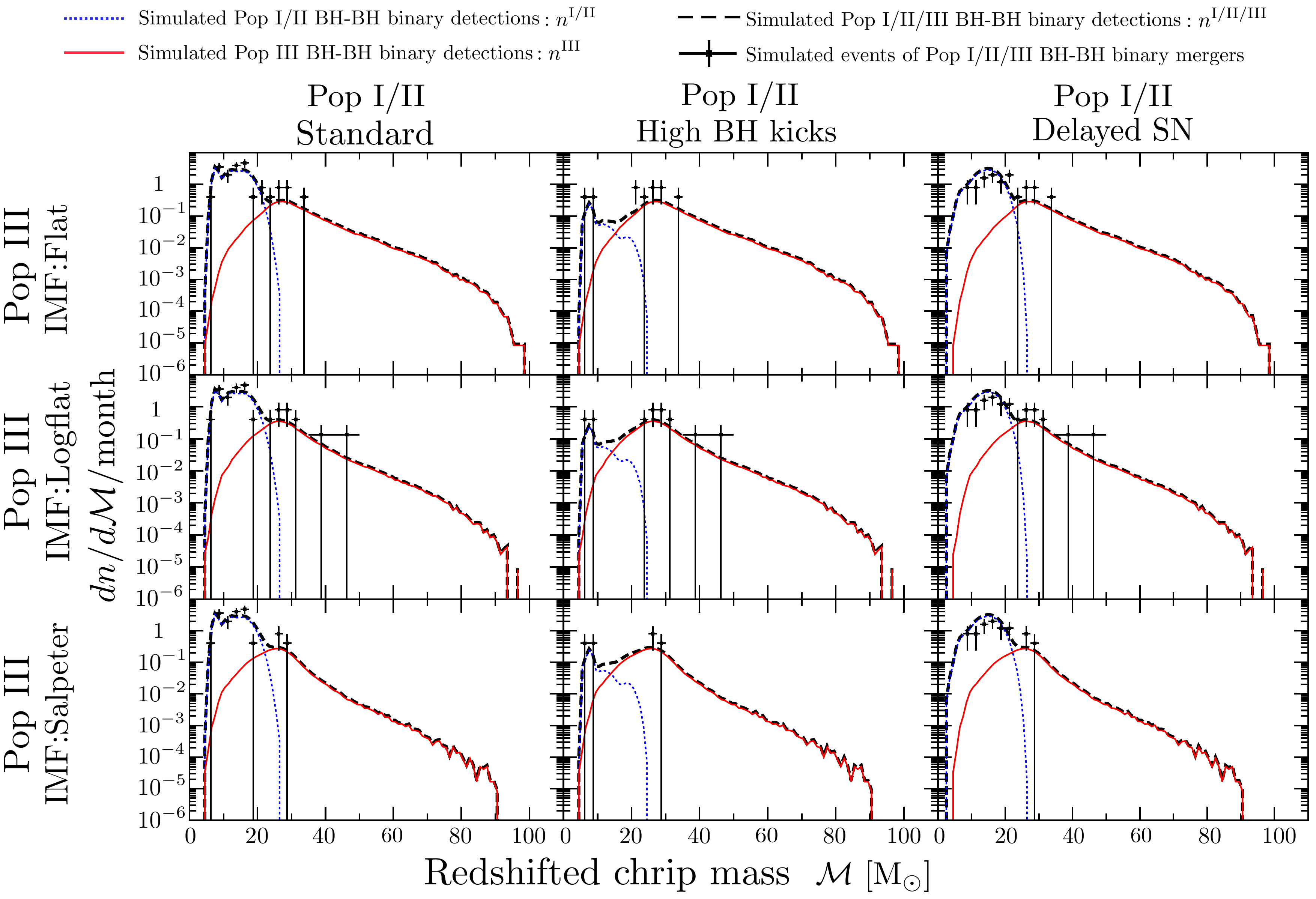}
	\caption{
Same as Fig.~\ref{f:nd-zmc-1model} but for more examples of population synthesis models.
From top to bottom panel, Pop III models are Kinugawa's IMF:Flat, IMF:Logflat, and IMF:Salpeter, respectively.
From left to right panel, Pop~I/II models are Dominik's standard model, high BH kicks model, and delayed SN model, respectively.
}
	\label{f:nd-zmc}
\end{figure*}

%%%%%%%%  Discussion and summary  %%%%%%%%
\section{Discussion and summary}
\label{sec:summary}
%%%%%%%%  Discussion  %%%%%%%%

%%%%%%%%  Population synthesis models  %%%%%%%%
\subsection{More population synthesis models}
\label{sec:moremodels}

%%  [Population synthesis, Pop I/II, Pop III] %%
We showed the result of the likelihood analysis with Dominik's standard model and Kinugawa's IMF:Flat as examples of models.
In this section, we show results of the likelihood analysis with more 2 examples of Pop~I/II models and more 2 examples of Pop~III models. So we show results of the likelihood analysis with $(1+2) \times (1+2) = 9$ combinations of Pop~I/II models and Pop~III models.

For more examples of Pop~I/II models, 
we employ three models of their models that are Dominik's standard model and models not used in previous sections.
The three models are the standard model and Domonik's variation 8 model submodel B (high BH kicks) and variation 10 model submodel B (delayed SN) with metallicity of $Z=\zsun$ and $Z=0.1~\zsun$ \cite{dominik2012}.
They use initial mass function $\Psi(m) \propto m^{-2.7}~(1.0~\msun \leq m < 150~\msun)$ for Pop~I/II model, where $m$ is the mass of the zero age main sequence star.
In the high BH kicks model model, BHs receive full natal kicks at the collapse. 
The delayed SN model uses ``delayed'' supernova model that produces an explosion as late as 1 second after the bounce, while the standard model uses the rapid supernova model.
In the case of delayed supernova model, a compact object mass distribution becomes a continuous distribution without BH mass gap \cite{belczynski2012}.
In their simulation, they calculated galactic merger rates of the local universe. To calculate merger rate densities within a few Gpc cubic, we employ the density of galaxies $\rho_\mathrm{gal} = 0.0116 ~\mathrm{Mpc}^{-3}$ \cite{kopparapu2008}.
We restrict Dominik's Pop~I/II BH-BH binaries that merge within 10 Gyrs.
These population synthesis data are also available on their online database.

For more Pop~III models, we prepare Kinugawa's the logflat IMF model (IMF:Logflat) and the Salpeter IMF model (IMF:Salpeter) with the following initial mass range and core-merger criterion during the common envelope phase.
The initial mass range of the Pop~III model is $10~\msun < m < 140~\msun$, where $m$ is a mass of the zero age main sequence star.
The initial mass function of IMF:Logflat and IMF:Salpeter are $\Psi(m) \propto m^{-1}$ and $\Psi(m) \propto m^{-2.35}$, respectively.
The core-merger criterion of the Pop~III model is a conservative one.

%%%%%%%%  Detection rate and mass distribution  %%%%%%%%
\subsection{Results with more examples of models}

For the more examples of population synthesis models, 
we estimated detection rates and average redshifted chirp mass distributions, and simulated observation results using the simulation of GW detections.
In Table~\ref{t:num-of-detect}, we show detection rates of examples of models.

In Fig.~\ref{f:nd-zmc}, we show average distributions of redshifted chirp mass $\mcz$ of
simulated BH-BH binary detections for a 1-month period, and simulated observation results for a 1-month period of GW detection in KAGRA.
The horizontal axis is the redshifted chirp mass. 
The vertical axis is the detection rate per redshifted chirp mass of BH-BH binary mergers [/$\msun$/month].
The dotted (blue) and solid (red) lines of Fig.~\ref{f:nd-zmc} are average distributions of redshifted chirp mass of simulated Pop~I/II BH-BH binary detections and Pop~III BH-BH binary detections for a 1-month period, respectively.
The dashed (black) line in Fig.~\ref{f:nd-zmc} corresponds to sum of the Pop~I/II model and the Pop~III model, i.e., Pop~I/II/III model.
The crosses and its error bars in Fig.~\ref{f:nd-zmc} show the simulated BH-BH binary detection events and its statistical errors on a 1-month period simulation, respectively.
Each panel of Fig.~\ref{f:nd-zmc} corresponds to each combination of Pop~I/II model and Pop~III model.

%%  [  Table I ]  %%
\begin{table}[h]
\caption{
Names of examples of the population synthesis models and estimated detection rates of BH-BH binary mergers for a 1-month period of observation in KAGRA.
}
\label{t:num-of-detect}
\centering
\begin{ruledtabular}
\begin{tabular}{lc}
Model name & Detection rate[/month]  \\ 
\hline
Pop~I/II Standard       &   31.99  \\     % 256521+3582193  = 3,838,714; --sqrt--> 1,959  (/10^4 yr)
Pop~I/II High BH kicks      &    1.050  \\     %       510+125473 = 125,983;     --sqrt-->  355
Pop~I/II Delayed SN     &   27.29  \\     % 5108+3269322  = 3,274,430   --sqrt-->  1,810
\hline
Pop~III \, IMF:Flat      &    4.685  \\     % 562205; --sqrt-->         750   (/10^4 yr)
Pop~III \, IMF:Logflat   &    5.182  \\     % 621884; --sqrt-->       789
Pop~III \, IMF:Salpeter  &    3.799       % 455845; --sqrt-->       675
\end{tabular}
\end{ruledtabular}
\end{table}

%%  [  Figure 5 ]  %%
\begin{figure*}
	\includegraphics[width=15cm]{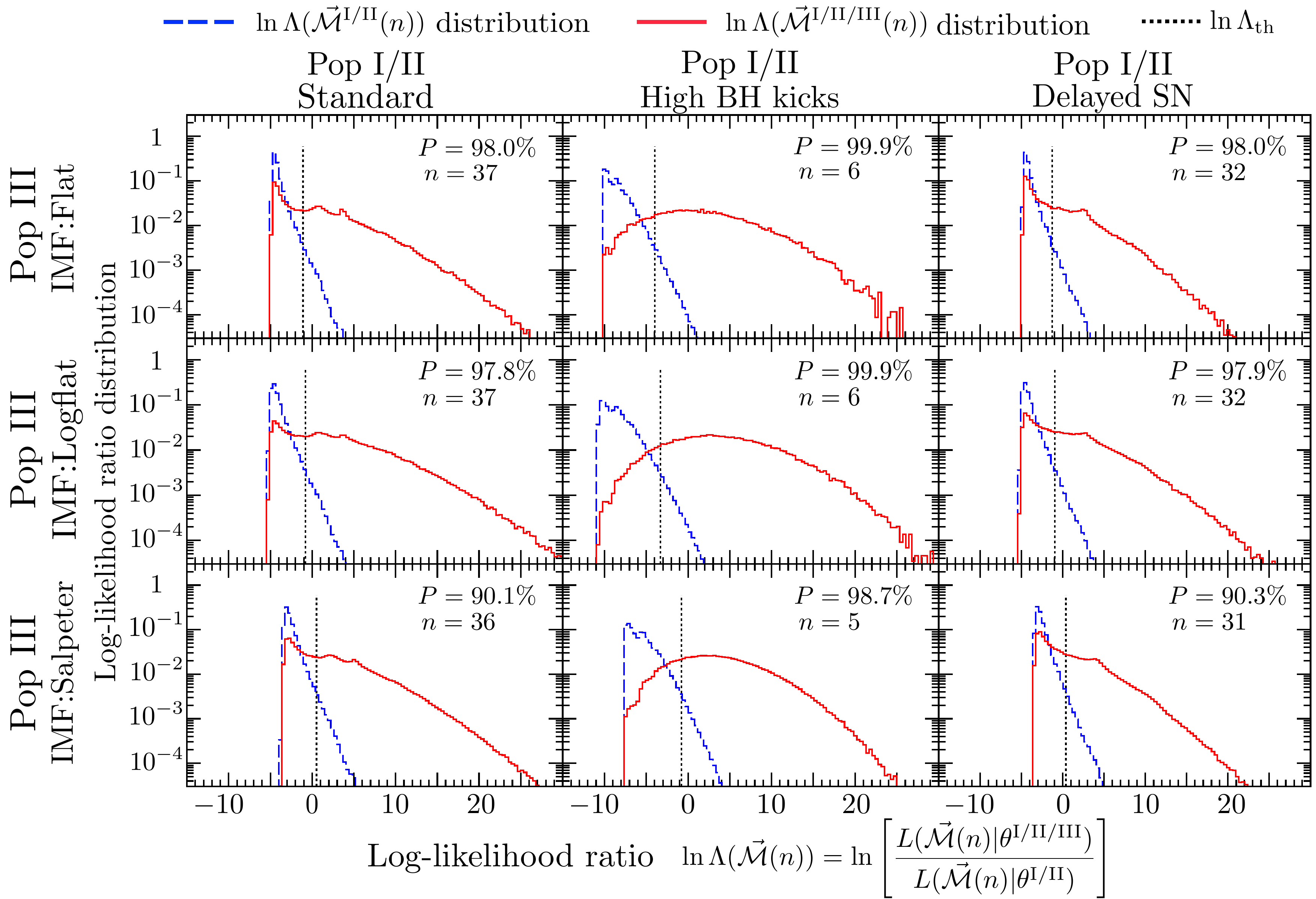}
	\caption{
Same as Figure~\ref{f:likelihood-hist-1mon-1model} but for more examples of population synthesis models.
From top to bottom panel, Pop III models are Kinugawa's IMF:Flat, IMF:Logflat, and IMF:Salpeter, respectively.
From left to right panel, Pop~I/II models are Dominik's standard model, high BH kicks model, and delayed SN model, respectively.
}
	\label{f:likelihood-hist-1mon}
\end{figure*}

%%%%%%%%  Evaluating probability that identifying the existence of Pop~III stars  %%%%%%%%
%\subsection{Evaluating probability that identifying the existence of Pop~III stars}

Next we show the probability of identifying the existence of Pop~III stars.
Figure~\ref{f:likelihood-hist-1mon} shows results of the likelihood analysis for equivalent a 1-month period of observation with more examples of models.
We use the detection rate of Pop~I/II/III BH-BH binary mergers rounded off to the nearest integer as the value of $n$.

In Figure~\ref{f:likelihood-hist-1mon}, we show the distributions of the log-likelihood ratio $\ln \Lambda(\vec{\bf{\mcz}}^\mathrm{I/II}(n))$ (dashed blue), the log-likelihood ratio thresholds $\ln \Lambda_\mathrm{th}$ (dotted black) and the distributions of the log-likelihood ratio $\ln \Lambda(\vec{\bf{\mcz}}^\mathrm{I/II/III}(n))$ (solid red)  for 9 combinations of models.

%%  [  Figure 6 ]  %%
\begin{figure}
\vspace{0.1cm}
	\includegraphics[width=8.5cm]{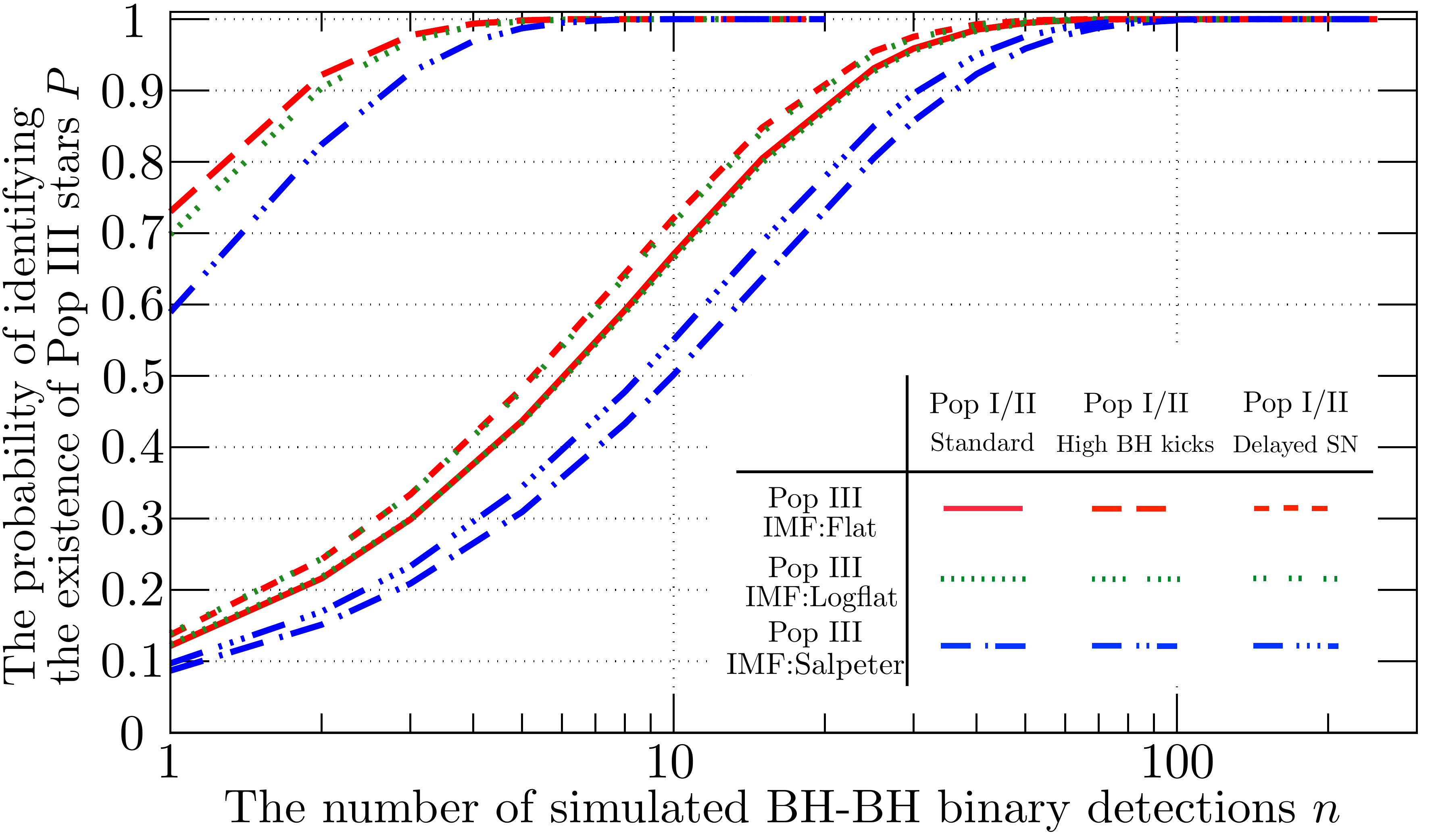}
\caption{
Same as Fig.~\ref{f:num-prob-P-1model} but for more examples of population synthesis models.
}
\label{f:num-prob-P}
\end{figure}

%%  [  Table II ]  %%
\begin{table*}
\small
\caption{
The probability of identifying the existence of Pop~III stars $P$.
The first column shows variations of Pop~I/II and Pop~III models. The second column shows the sum of the event rates of Pop~I/II model and Pop~III model.
The third column is the number of data sets $D_\mathrm{C}$ that Pop~I/II model and Pop~I/II/III model are clearly distinguished. 
The fourth column is the number of data sets $D_\mathrm{L}$ that Pop~I/II model and Pop~I/II/III model are distinguished by the likelihood analysis.
The fifth column shows the $P$ for equivalent a 1-month period of observation, respectively.
}
\label{t:events-probability}
\begin{ruledtabular}
  \begin{tabular}{lcrrc} 
   Pop~I/II ~~~~~~~ and Pop~III &  \shortstack{Event rate \\ ~[/month]}
   & $D_C$ 
   & $D_L$ 
   & $P = (D_C + D_L)/10^7$ \\ \hline 
  Standard ~~~~~~ and IMF:Flat           &     36.68   & 9 503 988 &  294 964 & 0.980 \\   
  Standard ~~~~~~ and IMF:Logflat      &     37.17     & 9 302 652 &  478 977 & 0.978 \\ 
  Standard ~~~~~~ and IMF:Salpeter   &     35.79       & 7 631 154 & 1382 485 & 0.901 \\
  High BH kicks and IMF:Flat           &     5.735  & 9 965 669 &   28 968 & 0.999 \\
  High BH kicks and IMF:Logflat      &     6.232    & 9 912 516 &   78 855 & 0.999 \\ 
  High BH kicks and IMF:Salpeter   &     4.849      & 9 333 580 &  538 701 & 0.987 \\
  Delayed SN ~~~ and IMF:Flat           &    31.98    & 9 628 499 &  176 166 & 0.980 \\ 
  Delayed SN ~~~ and IMF:Logflat      &    32.47      & 9 505 284 &  281 221 & 0.979 \\
  Delayed SN ~~~ and IMF:Salpeter   &    31.09        & 8 161 445 &  868 814 & 0.903   
\end{tabular}
 \end{ruledtabular}
\end{table*}

%%  [ probability of  {discrimination} as a function of #, Figure 3, model dependence, conclusion] %%
Next we show the dependence of the probability of identifying the existence of Pop~III stars $P$ on the number of simulated BH-BH binary detections $n$.
Figure~\ref{f:num-prob-P} shows the $P$ as a function of the number of simulated BH-BH binary detections.
The probability $P$ is growing with increasing the number of simulated BH-BH binary detections.
For standard model, the $P$ reaches 90\% at $n = 36$ events. 
For the high BH kicks model, the $P$ reaches 90\% at $n = 3$ events.
For the delayed SN model, the $P$ reaches 90\% at $n = 31$ events. 
Such number of simulated BH-BH binary detections $n$ can be achieved with a 1-month period observation.

%%  [rejected population synthesis model.]  %%
Here we compare the event rate of O1 run and the simulated event rates.
We also performed the simulation of GW detections from BH-BH binary mergers using Dominik's standard model submodel A of their paper \cite{dominik2012}. This model is also introduced in Ref.~\cite{dominik2013} as optimistic common envelope (optimistic CE) model. Then the detection rate of the model is 172.3 events/month in the design sensitivity of KAGRA. 
Assuming the effective observational duration of LIGO O1 run ($51.5$~days \cite{abbott2016b}) and the roughly $1/3$ times sensitivity of final KAGRA design, the expected number is approximately 
11 events.
This result is inconsistent with 2 events in O1 run.
Using the detection rates of Pop~I/II/III BH-BH binary mergers shown in Table~\ref{t:events-probability} are $4.8-37.2$ events/month, so we estimated that event rates of O1 run sensitivity is $0.3-2.4$ events/(51.5 days).  
Therefore these results are consistent with O1 run result.

%%%%%%%%  Analytical form of D_C  %%%%%%%%
\subsection{Analytical formula of $D_\mathrm{C}$}

%%  [model dependence of  {discrimination}]  %%
The $D_\mathrm{C}$ introduced in Sec.~\ref{sec:likelihood} can be described analytically.
In the case of $n=1$, $D_\mathrm{C}$ is described by
\begin{align}
D_\mathrm{C} = (1- q)D,
\label{e:PC-1event}
\end{align}
where 
\begin{align}
q  =  \int_{\mcz_{\min}^\mathrm{I/II}}^{\mcz_\mathrm{crit}^\mathrm{I/II} } p^\mathrm{I/II/III}(\mcz) ~d\mcz.
\label{e:q-definition}
\end{align}
 $q$ means a probability that the $\mcz$ of simulated BH-BH binary detections satisfies $\mcz_{\min}^\mathrm{I/II} < \mcz < \mcz_\mathrm{crit}^\mathrm{I/II}$ where $\mcz_{\min}^\mathrm{I/II}$ means minimum cutoff $\mcz$ of the detectable Pop~I/II BH-BH binary mergers.

In the case of arbitrary $n$, $D_\mathrm{C}$ is described by
\begin{align}
D_\mathrm{C} = (1- q^n)D.
\label{e:nC-n-event}
\end{align}
We summarize the $q$ in Table III. 
$D_\mathrm{C}$ depends on the probability $q$, the number of simulated BH-BH binary detections $n$ and the number of generated data sets $D$.

%%  [  Table III ]  %%
\begin{table}
\small
\caption{The probability $q$ defined in Eq.~(\ref{e:q-definition})}
\label{t:q}
\begin{ruledtabular}
  \begin{tabular}{cccc} 
                         &  Standard & High BH kicks & Delayed SN \\ \hline
  IMF:Flat          &      0.9223791 & 0.3896844 & 0.9024576 \\
  IMF:Logflat      &       0.9309342 & 0.4556952 & 0.9106011 \\ 
  IMF:Salpeter   &         0.9610642 & 0.5828035 & 0.9471150 \\
  \end{tabular}
 \end{ruledtabular}
\end{table}

%%%%%%%%  Dependence of results on parameters of the population synthesis model  %%%%%%%%
\subsection{Dependence of results on parameters of the population synthesis model}
We discuss dependence of our results on parameters of the population synthesis model.
Our results depend on binary parameters the star formation rate (SFR).
In the case of Pop~I/II, the SFR is well known by the observation, but the merger rate of Pop~I/II BH-BH binary depends heavily on the binary parameters, especially the common envelope parameters $\alpha'$ and $\lambda$. $\alpha'$ is the efficiency of energy transfer and $\lambda$ is the parameter of the binding energy of primary envelope.
When the secondary plunges into the envelope of primary giant, the binary becomes the common envelope phase. During the common envelope phase, the friction between the secondary and the envelope dissipates the orbital energy and the secondary spirals in.
The envelope is evaporated because the lost orbital energy is used to eject the envelope.
Finally, the binary becomes the close binary which consists of the secondary and the core of primary giant or merges.
If $\alpha' \lambda$ change, the merger rate of Pop~I/II BH-BH binary changes by $2-3$ orders of magnitude \cite{Madau2014} due to almost BH-BH binary progenitors of Pop~I/II evolve via a common envelope phase \cite{Ivanova2012}.
Recently, $\lambda$ is calculated by the stellar evolutions \cite{Xu2010} and the population synthesis calculations use the realistic $\lambda$ \cite{Inayoshi2017}.
However, the uncertainty of $\alpha'$ remains yet and the merger rate is changed by the uncertainty of $\alpha'$.
On the other hand, in the case of Pop~III, the merger rate is only changed a few times by the binary parameters because the BH-BH binary progenitors of Pop~III can evolve without a common envelope phase \cite{Visbal2015}.
However, the SFR of Pop~III has large uncertainty.
Recently, researchers try to constraint the SFR of Pop~III by the optical depth of Thomson scattering observed by Planck \cite{Ade2015} although there are some uncertainty parameters \cite{hartwig2016, Inayoshi2016}.
According to these results, the SFR of Pop~III might be $3-10$ times smaller than that of our calculation.

Next, we discuss the IMF of Pop~II and the effect of its uncertainty on our results.
Even if Pop~II stars have flat IMF, the peak of chirp mass distribution of Pop~II BH-BH binaries is different from one of Pop~III BH-BH binaries because the evolution of Pop~II stars is different from one of Pop~III stars.
The evolution of Pop~III stars differs depending on whether a mass of zero age main sequence star is larger or smaller than $50~\msun$. Finally the chirp mass distribution of Pop~III BH-BH binaries has a peak at $30~\msun$.
On the other hand, since Pop~II stars evolve uniformly as red giants, the evolution path is uniform even if the mass changes. Therefore, it seems that the chirp mass distribution of Pop~II BH-BH binaries does not have a characteristic peak and it distributes to trace the IMF of Pop~II.
Therefore, if the IMF of Pop~II is flat, the number of high mass Pop~II BH-BH binaries will increase, so it will be difficult to find the Pop~III BH-BH binaries. However, since the positions of the peaks of the chirp mass distribution of Pop~II and Pop~III BH-BH binaries do not overlap, the existence of Pop~III stars can be confirmed if there are many Pop~III BH-BH binaries.
In addition, it is known from the calculation of star formation that the star-formation clouds split with lower mass \cite{Omukai2005}, if there is metal in the clouds.
So the IMF of Pop~II seems to be less likely to be the IMF of Pop~III with many massive stars such as the flat IMF.

The number of data sets that Pop~I/II model and Pop~I/II/III model are clearly distinguished $D_\mathrm{C}$
depends on $\mcz_\mathrm{crit}^\mathrm{I/II}$, the number of Pop~III BH-BH binary detections and the number of generated data sets $D$.
$\mcz_\mathrm{crit}^\mathrm{I/II}$ depends on the binary parameters of Pop~I/II model.
If $\mcz_\mathrm{crit}^\mathrm{I/II}$ is larger than $\mcz_\mathrm{crit}^\mathrm{I/II}$ of Pop~I/II model which we employed, $D_\mathrm{C}$ will become smaller than our results.
The number of Pop~III BH-BH binary detections depends on SFR of Pop~III.
The smaller SFR of Pop~III, the smaller the number of Pop~III BH-BH binary detections.
The smaller the number of Pop~III BH-BH binary detections, the smaller $D_\mathrm{C}$.

The number of data sets that Pop~I/II model and Pop~I/II/III model are distinguished by the likelihood analysis $D_\mathrm{L}$ depends on the number of Pop~I/II BH-BH binary detections, the number of Pop~III BH-BH binary detections and the number of generated data sets $D$.
The number of Pop~I/II BH-BH binary detections depends on the binary parameters of Pop~I/II model.
The number of Pop~III BH-BH binary detections depends on the SFR of Pop~III model.
If the ratio of the number of BH-BH binary detections of Pop~III to Pop~I/II is larger than Pop~I/II/III model which we employed, $D_\mathrm{L}$ will be larger than our results.

%%%%%%%%  Another way to identify the existence of Pop~III stars  %%%%%%%%
\subsection{Other ways to identify the existence of Pop~III stars}

We performed the likelihood analysis of redshifted chirp mass distributions of the population synthesis models.
We evaluated the probability of identifying the existence of Pop~III stars.
The likelihood analysis using information of BH-BH binary mergers could become a way to distinguish population synthesis models.

There are other ways to identify the existence of Pop~III stars.
(i) Inayoshi {\it et al.} \cite{Inayoshi2016} proposed that the stochastic background from a Pop~III star contribution will deviate measurably from the canonical ~2/3rd slope of $d \ln \Omega_\mathrm{gw}/d\ln f$, where $\Omega_\mathrm{gw}$ is the amplitude of GW background.
(ii) Pacucci {\it et al.} \cite{Pacucci2017} proposed that Pop~III stellar remnant BHs are located preferentially in the bulges of galaxies and if their location can be measured, this would be evidence for their Pop~III origin.
(iii) Inayoshi {\it et al.} \cite{Inayoshi2017arxiv} proposed that if Pop~III stellar remnant BHs are indeed located preferentially in the bulges, and within the sphere of influence of the central supermassive BH, then the center-of-mass acceleration of the binary can be measured (with LISA \cite{lisa} + LIGO).
(iv) Pop~I/II and Pop~III BH-BH binary mergers have the different redshift dependence. 
In the Pop~I/II model, the cumulative event rate saturates at $z \sim 6$.  
On the other hand, in the Pop~III model, the cumulative event rate saturates at $z \sim 10$ \cite{decigo}.  
Since the Pop III stars are obviously formed earlier than Pop I/II stars, it is robust that there is the difference of redshift dependence.
Since the detectable range of a second generation terrestrial GW detector is up to $z\sim2$, it cannot measure the difference of redshift dependence.
However, the space GW antennae such as B-DECIGO \cite{decigo} and LISA \cite{lisa} can detect BH-BH binary mergers at $z \sim 30$ and measure the redshift dependence.
We would be able to identify the existence of Pop~III stars using the luminosity distance $d_\mathrm{L}$ or redshift.
Furthermore it will be able to distinguish in detail between population synthesis models by performing our likelihood analysis using BH-BH binary detections at each redshift.

%%%%%%%%  Acknowledgments  %%%%%%%%
\begin{acknowledgments}
This work was supported by MEXT Grant-in-Aid for Scientific Research on Innovative Areas "New Developments in Astrophysics Through Multi-Messenger Observations of Gravitational Wave Sources" (Grant Number 24103005, 24103006), and the Grant-in-Aid from the Ministry of Education, Culture, Sports, Science and Technology (MEXT) of Japan Grant No. 15H02087 (TN) and JSPS KAKENHI Grant No. JP16818962 (TK).
% {, 23540305, 15H02087), JSPS Leading-edge Research Infrastructure Program (251284), JSPS Grant-in-Aid for Specially Promoted Research 26000005,  JSPS Core-to-Core Program, A. Advanced Research Networks, and the joint research program of the Institute for Cosmic Ray Research, University of Tokyo.}
\end{acknowledgments}

%%%%%%%%  References  %%%%%%%%
\bibliography{references}

\begin{thebibliography}{37}%
\makeatletter
\providecommand \@ifxundefined [1]{%
 \@ifx{#1\undefined}
}%
\providecommand \@ifnum [1]{%
 \ifnum #1\expandafter \@firstoftwo
 \else \expandafter \@secondoftwo
 \fi
}%
\providecommand \@ifx [1]{%
 \ifx #1\expandafter \@firstoftwo
 \else \expandafter \@secondoftwo
 \fi
}%
\providecommand \natexlab [1]{#1}%
\providecommand \enquote  [1]{``#1''}%
\providecommand \bibnamefont  [1]{#1}%
\providecommand \bibfnamefont [1]{#1}%
\providecommand \citenamefont [1]{#1}%
\providecommand \href@noop [0]{\@secondoftwo}%
\providecommand \href [0]{\begingroup \@sanitize@url \@href}%
\providecommand \@href[1]{\@@startlink{#1}\@@href}%
\providecommand \@@href[1]{\endgroup#1\@@endlink}%
\providecommand \@sanitize@url [0]{\catcode `\\12\catcode `\$12\catcode
  `\&12\catcode `\#12\catcode `\^12\catcode `\_12\catcode `\%12\relax}%
\providecommand \@@startlink[1]{}%
\providecommand \@@endlink[0]{}%
\providecommand \url  [0]{\begingroup\@sanitize@url \@url }%
\providecommand \@url [1]{\endgroup\@href {#1}{\urlprefix }}%
\providecommand \urlprefix  [0]{URL }%
\providecommand \Eprint [0]{\href }%
\providecommand \doibase [0]{http://dx.doi.org/}%
\providecommand \selectlanguage [0]{\@gobble}%
\providecommand \bibinfo  [0]{\@secondoftwo}%
\providecommand \bibfield  [0]{\@secondoftwo}%
\providecommand \translation [1]{[#1]}%
\providecommand \BibitemOpen [0]{}%
\providecommand \bibitemStop [0]{}%
\providecommand \bibitemNoStop [0]{.\EOS\space}%
\providecommand \EOS [0]{\spacefactor3000\relax}%
\providecommand \BibitemShut  [1]{\csname bibitem#1\endcsname}%
\let\auto@bib@innerbib\@empty
%</preamble>
\bibitem [{\citenamefont {Harry}(2010)}]{LIGOpaper}%
  \BibitemOpen
  \bibfield  {author} {\bibinfo {author} {\bibfnamefont {G.~M.}\ \bibnamefont
  {Harry}} (\bibinfo {collaboration} {LIGO Scientific Collaboration}),\
  }\href@noop {} {\bibfield  {journal} {\bibinfo  {journal} {Classical Quantum
  Gravity}\ }\textbf {\bibinfo {volume} {27}},\ \bibinfo {pages} {084006}
  (\bibinfo {year} {2010})}\BibitemShut {NoStop}%
\bibitem [{\citenamefont {Abbott}\ \emph
  {et~al.}(2016{\natexlab{a}})\citenamefont {Abbott} \emph
  {et~al.}}]{abbott2016}%
  \BibitemOpen
  \bibfield  {author} {\bibinfo {author} {\bibfnamefont {B.~P.}\ \bibnamefont
  {Abbott}} \emph {et~al.} (\bibinfo {collaboration} {LIGO Scientific
  Collaboration and Virgo Collaboration}),\ }\href@noop {} {\bibfield
  {journal} {\bibinfo  {journal} {Phys. Rev. Lett.}\ }\textbf {\bibinfo
  {volume} {116}},\ \bibinfo {pages} {061102} (\bibinfo {year}
  {2016}{\natexlab{a}})}\BibitemShut {NoStop}%
\bibitem [{\citenamefont {Kinugawa}\ \emph {et~al.}(2014)\citenamefont
  {Kinugawa}, \citenamefont {Inayoshi}, \citenamefont {Hotokezaka},
  \citenamefont {Nakauchi},\ and\ \citenamefont {Nakamura}}]{kinugawa2014}%
  \BibitemOpen
  \bibfield  {author} {\bibinfo {author} {\bibfnamefont {T.}~\bibnamefont
  {Kinugawa}}, \bibinfo {author} {\bibfnamefont {K.}~\bibnamefont {Inayoshi}},
  \bibinfo {author} {\bibfnamefont {K.}~\bibnamefont {Hotokezaka}}, \bibinfo
  {author} {\bibfnamefont {D.}~\bibnamefont {Nakauchi}}, \ and\ \bibinfo
  {author} {\bibfnamefont {T.}~\bibnamefont {Nakamura}},\ }\href@noop {}
  {\bibfield  {journal} {\bibinfo  {journal} {Mon. Not. R. Astron. Soc.}\
  }\textbf {\bibinfo {volume} {442}},\ \bibinfo {pages} {2963} (\bibinfo {year}
  {2014})}\BibitemShut {NoStop}%
\bibitem [{\citenamefont {Kinugawa}\ \emph {et~al.}(2016)\citenamefont
  {Kinugawa}, \citenamefont {Miyamoto}, \citenamefont {Kanda},\ and\
  \citenamefont {Nakamura}}]{kinugawa2016}%
  \BibitemOpen
  \bibfield  {author} {\bibinfo {author} {\bibfnamefont {T.}~\bibnamefont
  {Kinugawa}}, \bibinfo {author} {\bibfnamefont {A.}~\bibnamefont {Miyamoto}},
  \bibinfo {author} {\bibfnamefont {N.}~\bibnamefont {Kanda}}, \ and\ \bibinfo
  {author} {\bibfnamefont {T.}~\bibnamefont {Nakamura}},\ }\href@noop {}
  {\bibfield  {journal} {\bibinfo  {journal} {Mon. Not. R. Astron. Soc.}\
  }\textbf {\bibinfo {volume} {456}},\ \bibinfo {pages} {1093} (\bibinfo {year}
  {2016})}\BibitemShut {NoStop}%
\bibitem [{\citenamefont {Acernese}\ \emph {et~al.}(2015)\citenamefont
  {Acernese} \emph {et~al.}}]{Virgopaper}%
  \BibitemOpen
  \bibfield  {author} {\bibinfo {author} {\bibfnamefont {F.}~\bibnamefont
  {Acernese}} \emph {et~al.} (\bibinfo {collaboration} {Virgo Collaboration}),\
  }\href@noop {} {\bibfield  {journal} {\bibinfo  {journal} {Classical Quantum
  Gravity}\ }\textbf {\bibinfo {volume} {32}},\ \bibinfo {pages} {024001}
  (\bibinfo {year} {2015})}\BibitemShut {NoStop}%
\bibitem [{\citenamefont {KAGRA}()}]{kagra}%
  \BibitemOpen
  \bibfield  {author} {\bibinfo {author} {\bibnamefont {KAGRA}},\ }\href@noop
  {} {}\bibinfo {howpublished}
  {\url{http://gwcenter.icrr.u-tokyo.ac.jp/en/}}\BibitemShut {NoStop}%
\bibitem [{\citenamefont {Nakamura}\ \emph {et~al.}(1997)\citenamefont
  {Nakamura}, \citenamefont {Sasaki}, \citenamefont {Tanaka},\ and\
  \citenamefont {Thorne}}]{nakamura1997}%
  \BibitemOpen
  \bibfield  {author} {\bibinfo {author} {\bibfnamefont {T.}~\bibnamefont
  {Nakamura}}, \bibinfo {author} {\bibfnamefont {M.}~\bibnamefont {Sasaki}},
  \bibinfo {author} {\bibfnamefont {T.}~\bibnamefont {Tanaka}}, \ and\ \bibinfo
  {author} {\bibfnamefont {K.~S.}\ \bibnamefont {Thorne}},\ }\href@noop {}
  {\bibfield  {journal} {\bibinfo  {journal} {Astrophys. J. Lett.}\ }\textbf
  {\bibinfo {volume} {487}},\ \bibinfo {pages} {L139} (\bibinfo {year}
  {1997})}\BibitemShut {NoStop}%
\bibitem [{\citenamefont {Sasaki}\ \emph {et~al.}(2016)\citenamefont {Sasaki},
  \citenamefont {Suyama}, \citenamefont {Tanaka},\ and\ \citenamefont
  {Yokoyama}}]{sasaki2016}%
  \BibitemOpen
  \bibfield  {author} {\bibinfo {author} {\bibfnamefont {M.}~\bibnamefont
  {Sasaki}}, \bibinfo {author} {\bibfnamefont {T.}~\bibnamefont {Suyama}},
  \bibinfo {author} {\bibfnamefont {T.}~\bibnamefont {Tanaka}}, \ and\ \bibinfo
  {author} {\bibfnamefont {S.}~\bibnamefont {Yokoyama}},\ }\href@noop {}
  {\bibfield  {journal} {\bibinfo  {journal} {Phys. Rev. Lett.}\ }\textbf
  {\bibinfo {volume} {117}},\ \bibinfo {pages} {061101} (\bibinfo {year}
  {2016})}\BibitemShut {NoStop}%
\bibitem [{\citenamefont {{Portegies Zwart}}\ and\ \citenamefont
  {McMillan}(2000)}]{portegies2000}%
  \BibitemOpen
  \bibfield  {author} {\bibinfo {author} {\bibfnamefont {S.~F.}\ \bibnamefont
  {{Portegies Zwart}}}\ and\ \bibinfo {author} {\bibfnamefont {S.~L.~W.}\
  \bibnamefont {McMillan}},\ }\href@noop {} {\bibfield  {journal} {\bibinfo
  {journal} {Astrophys. J. Lett.}\ }\textbf {\bibinfo {volume} {528}},\
  \bibinfo {pages} {L17} (\bibinfo {year} {2000})}\BibitemShut {NoStop}%
\bibitem [{\citenamefont {Zevin}\ \emph {et~al.}()\citenamefont {Zevin},
  \citenamefont {Pankow}, \citenamefont {Rodriguez}, \citenamefont {Sampson},
  \citenamefont {Chase}, \citenamefont {Kalogera},\ and\ \citenamefont
  {Rasio}}]{zevin2017}%
  \BibitemOpen
  \bibfield  {author} {\bibinfo {author} {\bibfnamefont {M.}~\bibnamefont
  {Zevin}}, \bibinfo {author} {\bibfnamefont {C.}~\bibnamefont {Pankow}},
  \bibinfo {author} {\bibfnamefont {C.~L.}\ \bibnamefont {Rodriguez}}, \bibinfo
  {author} {\bibfnamefont {L.}~\bibnamefont {Sampson}}, \bibinfo {author}
  {\bibfnamefont {E.}~\bibnamefont {Chase}}, \bibinfo {author} {\bibfnamefont
  {V.}~\bibnamefont {Kalogera}}, \ and\ \bibinfo {author} {\bibfnamefont
  {F.~A.}\ \bibnamefont {Rasio}},\ }\href@noop {} {\ }\Eprint
  {http://arxiv.org/abs/1704.07379} {arXiv:1704.07379} \BibitemShut {NoStop}%
\bibitem [{\citenamefont {Dominik}\ \emph {et~al.}(2012)\citenamefont
  {Dominik}, \citenamefont {Belczynski}, \citenamefont {Fryer}, \citenamefont
  {Holz}, \citenamefont {Berti}, \citenamefont {Bulik}, \citenamefont
  {Mandel},\ and\ \citenamefont {O'Shaughnessy}}]{dominik2012}%
  \BibitemOpen
  \bibfield  {author} {\bibinfo {author} {\bibfnamefont {M.}~\bibnamefont
  {Dominik}}, \bibinfo {author} {\bibfnamefont {K.}~\bibnamefont {Belczynski}},
  \bibinfo {author} {\bibfnamefont {C.}~\bibnamefont {Fryer}}, \bibinfo
  {author} {\bibfnamefont {D.~E.}\ \bibnamefont {Holz}}, \bibinfo {author}
  {\bibfnamefont {E.}~\bibnamefont {Berti}}, \bibinfo {author} {\bibfnamefont
  {T.}~\bibnamefont {Bulik}}, \bibinfo {author} {\bibfnamefont
  {I.}~\bibnamefont {Mandel}}, \ and\ \bibinfo {author} {\bibfnamefont
  {R.}~\bibnamefont {O'Shaughnessy}},\ }\href@noop {} {\bibfield  {journal}
  {\bibinfo  {journal} {Astrophys. J.}\ }\textbf {\bibinfo {volume} {759}},\
  \bibinfo {pages} {52} (\bibinfo {year} {2012})}\BibitemShut {NoStop}%
\bibitem [{\citenamefont {Dominik}\ \emph {et~al.}(2013)\citenamefont
  {Dominik}, \citenamefont {Belczynski}, \citenamefont {Fryer}, \citenamefont
  {Holz}, \citenamefont {Berti}, \citenamefont {Bulik}, \citenamefont
  {Mandel},\ and\ \citenamefont {O'Shaughnessy}}]{dominik2013}%
  \BibitemOpen
  \bibfield  {author} {\bibinfo {author} {\bibfnamefont {M.}~\bibnamefont
  {Dominik}}, \bibinfo {author} {\bibfnamefont {K.}~\bibnamefont {Belczynski}},
  \bibinfo {author} {\bibfnamefont {C.}~\bibnamefont {Fryer}}, \bibinfo
  {author} {\bibfnamefont {D.~E.}\ \bibnamefont {Holz}}, \bibinfo {author}
  {\bibfnamefont {E.}~\bibnamefont {Berti}}, \bibinfo {author} {\bibfnamefont
  {T.}~\bibnamefont {Bulik}}, \bibinfo {author} {\bibfnamefont
  {I.}~\bibnamefont {Mandel}}, \ and\ \bibinfo {author} {\bibfnamefont
  {R.}~\bibnamefont {O'Shaughnessy}},\ }\href@noop {} {\bibfield  {journal}
  {\bibinfo  {journal} {Astrophys. J.}\ }\textbf {\bibinfo {volume} {779}},\
  \bibinfo {pages} {72} (\bibinfo {year} {2013})}\BibitemShut {NoStop}%
\bibitem [{\citenamefont {Kopparapu}\ \emph {et~al.}(2008)\citenamefont
  {Kopparapu}, \citenamefont {Hanna}, \citenamefont {Kalogera}, \citenamefont
  {O’Shaughnessy}, \citenamefont {Gonz\'{a}lez}, \citenamefont {Brady},\ and\
  \citenamefont {Fairhurst}}]{kopparapu2008}%
  \BibitemOpen
  \bibfield  {author} {\bibinfo {author} {\bibfnamefont {R.~K.}\ \bibnamefont
  {Kopparapu}}, \bibinfo {author} {\bibfnamefont {C.}~\bibnamefont {Hanna}},
  \bibinfo {author} {\bibfnamefont {V.}~\bibnamefont {Kalogera}}, \bibinfo
  {author} {\bibfnamefont {R.}~\bibnamefont {O’Shaughnessy}}, \bibinfo
  {author} {\bibfnamefont {G.}~\bibnamefont {Gonz\'{a}lez}}, \bibinfo {author}
  {\bibfnamefont {P.~R.}\ \bibnamefont {Brady}}, \ and\ \bibinfo {author}
  {\bibfnamefont {S.}~\bibnamefont {Fairhurst}},\ }\href@noop {} {\bibfield
  {journal} {\bibinfo  {journal} {Astrophys. J.}\ }\textbf {\bibinfo {volume}
  {675}},\ \bibinfo {pages} {1459} (\bibinfo {year} {2008})}\BibitemShut
  {NoStop}%
\bibitem [{\citenamefont {Belczynski}\ \emph {et~al.}(2002)\citenamefont
  {Belczynski}, \citenamefont {Kalogera},\ and\ \citenamefont
  {Bulik}}]{belczynski2002}%
  \BibitemOpen
  \bibfield  {author} {\bibinfo {author} {\bibfnamefont {K.}~\bibnamefont
  {Belczynski}}, \bibinfo {author} {\bibfnamefont {V.}~\bibnamefont
  {Kalogera}}, \ and\ \bibinfo {author} {\bibfnamefont {T.}~\bibnamefont
  {Bulik}},\ }\href@noop {} {\bibfield  {journal} {\bibinfo  {journal}
  {Astrophys. J.}\ }\textbf {\bibinfo {volume} {572}},\ \bibinfo {pages} {407}
  (\bibinfo {year} {2002})}\BibitemShut {NoStop}%
\bibitem [{\citenamefont {Spergel}\ \emph {et~al.}(2007)\citenamefont {Spergel}
  \emph {et~al.}}]{spergel2007}%
  \BibitemOpen
  \bibfield  {author} {\bibinfo {author} {\bibfnamefont {D.~N.}\ \bibnamefont
  {Spergel}} \emph {et~al.},\ }\href@noop {} {\bibfield  {journal} {\bibinfo
  {journal} {Astrophys. J. Suppl. Ser.}\ }\textbf {\bibinfo {volume} {170}},\
  \bibinfo {pages} {377} (\bibinfo {year} {2007})}\BibitemShut {NoStop}%
\bibitem [{\citenamefont {Ade}\ \emph {et~al.}(2014)\citenamefont {Ade} \emph
  {et~al.}}]{planck2013}%
  \BibitemOpen
  \bibfield  {author} {\bibinfo {author} {\bibfnamefont {P.~A.~R.}\
  \bibnamefont {Ade}} \emph {et~al.} (\bibinfo {collaboration} {Planck
  Collaboration}),\ }\href@noop {} {\bibfield  {journal} {\bibinfo  {journal}
  {Astron. Astrophys.}\ }\textbf {\bibinfo {volume} {571}},\ \bibinfo {pages}
  {A16} (\bibinfo {year} {2014})}\BibitemShut {NoStop}%
\bibitem [{\citenamefont {Blanchet}\ \emph {et~al.}(1996)\citenamefont
  {Blanchet}, \citenamefont {Iyer}, \citenamefont {Will},\ and\ \citenamefont
  {Wiseman}}]{blanchet1996}%
  \BibitemOpen
  \bibfield  {author} {\bibinfo {author} {\bibfnamefont {L.}~\bibnamefont
  {Blanchet}}, \bibinfo {author} {\bibfnamefont {B.~R.}\ \bibnamefont {Iyer}},
  \bibinfo {author} {\bibfnamefont {C.~M.}\ \bibnamefont {Will}}, \ and\
  \bibinfo {author} {\bibfnamefont {A.~G.}\ \bibnamefont {Wiseman}},\
  }\href@noop {} {\bibfield  {journal} {\bibinfo  {journal} {Classical Quantum
  Gravity}\ }\textbf {\bibinfo {volume} {13}},\ \bibinfo {pages} {575}
  (\bibinfo {year} {1996})}\BibitemShut {NoStop}%
\bibitem [{\citenamefont {Sathyaprakash}\ and\ \citenamefont
  {Schutz}(2009)}]{sathyaprakash2009}%
  \BibitemOpen
  \bibfield  {author} {\bibinfo {author} {\bibfnamefont {B.~S.}\ \bibnamefont
  {Sathyaprakash}}\ and\ \bibinfo {author} {\bibfnamefont {B.~F.}\ \bibnamefont
  {Schutz}},\ }\href@noop {} {\bibfield  {journal} {\bibinfo  {journal} {Living
  Rev. Relativ.}\ }\textbf {\bibinfo {volume} {12}},\ \bibinfo {pages} {2}
  (\bibinfo {year} {2009})}\BibitemShut {NoStop}%
\bibitem [{\citenamefont {Abbott}\ \emph
  {et~al.}(2016{\natexlab{b}})\citenamefont {Abbott} \emph
  {et~al.}}]{propertygw150914}%
  \BibitemOpen
  \bibfield  {author} {\bibinfo {author} {\bibfnamefont {B.~P.}\ \bibnamefont
  {Abbott}} \emph {et~al.} (\bibinfo {collaboration} {LIGO Scientific
  Collaboration and Virgo Collaboration}),\ }\href@noop {} {\bibfield
  {journal} {\bibinfo  {journal} {Phys. Rev. Lett.}\ }\textbf {\bibinfo
  {volume} {116}},\ \bibinfo {pages} {241102} (\bibinfo {year}
  {2016}{\natexlab{b}})}\BibitemShut {NoStop}%
\bibitem [{\citenamefont {Abbott}\ \emph
  {et~al.}(2016{\natexlab{c}})\citenamefont {Abbott} \emph {et~al.}}]{O1run}%
  \BibitemOpen
  \bibfield  {author} {\bibinfo {author} {\bibfnamefont {B.~P.}\ \bibnamefont
  {Abbott}} \emph {et~al.} (\bibinfo {collaboration} {LIGO Scientific
  Collaboration and Virgo Collaboration}),\ }\href@noop {} {\bibfield
  {journal} {\bibinfo  {journal} {Phys. Rev. X}\ }\textbf {\bibinfo {volume}
  {6}},\ \bibinfo {pages} {041015} (\bibinfo {year}
  {2016}{\natexlab{c}})}\BibitemShut {NoStop}%
\bibitem [{\citenamefont {Flanagan}\ and\ \citenamefont
  {Hughes}(1998)}]{Flanagan1998}%
  \BibitemOpen
  \bibfield  {author} {\bibinfo {author} {\bibfnamefont {{\'E}.~{\'E}.}\
  \bibnamefont {Flanagan}}\ and\ \bibinfo {author} {\bibfnamefont {S.~A.}\
  \bibnamefont {Hughes}},\ }\href@noop {} {\bibfield  {journal} {\bibinfo
  {journal} {Phys. Rev. D}\ }\textbf {\bibinfo {volume} {57}},\ \bibinfo
  {pages} {4535} (\bibinfo {year} {1998})}\BibitemShut {NoStop}%
\bibitem [{\citenamefont {Creighton}\ and\ \citenamefont
  {Anderson}(2011)}]{creightontext}%
  \BibitemOpen
  \bibfield  {author} {\bibinfo {author} {\bibfnamefont {J.}~\bibnamefont
  {Creighton}}\ and\ \bibinfo {author} {\bibfnamefont {W.}~\bibnamefont
  {Anderson}},\ }\href@noop {} {\emph {\bibinfo {title} {Gravitational-Wave
  Physics and Astronomy: An Introduction to Theory, Experiment and Data
  Analysis}}},\ Wiley Series in Cosmology\ (\bibinfo  {publisher} {Wiley, New
  York},\ \bibinfo {year} {2011})\BibitemShut {NoStop}%
\bibitem [{\citenamefont {Belczynski}\ \emph {et~al.}(2012)\citenamefont
  {Belczynski}, \citenamefont {Wiktorowicz}, \citenamefont {Fryer},
  \citenamefont {Holz},\ and\ \citenamefont {Kalogera}}]{belczynski2012}%
  \BibitemOpen
  \bibfield  {author} {\bibinfo {author} {\bibfnamefont {K.}~\bibnamefont
  {Belczynski}}, \bibinfo {author} {\bibfnamefont {G.}~\bibnamefont
  {Wiktorowicz}}, \bibinfo {author} {\bibfnamefont {C.}~\bibnamefont {Fryer}},
  \bibinfo {author} {\bibfnamefont {D.~E.}\ \bibnamefont {Holz}}, \ and\
  \bibinfo {author} {\bibfnamefont {V.}~\bibnamefont {Kalogera}},\ }\href@noop
  {} {\bibfield  {journal} {\bibinfo  {journal} {Astrophys. J.}\ }\textbf
  {\bibinfo {volume} {757}},\ \bibinfo {pages} {91} (\bibinfo {year}
  {2012})}\BibitemShut {NoStop}%
\bibitem [{\citenamefont {Abbott}\ \emph
  {et~al.}(2016{\natexlab{d}})\citenamefont {Abbott} \emph
  {et~al.}}]{abbott2016b}%
  \BibitemOpen
  \bibfield  {author} {\bibinfo {author} {\bibfnamefont {B.~P.}\ \bibnamefont
  {Abbott}} \emph {et~al.} (\bibinfo {collaboration} {LIGO Scientific
  Collaboration and Virgo Collaboration}),\ }\href@noop {} {\bibfield
  {journal} {\bibinfo  {journal} {Phys. Rev. X}\ }\textbf {\bibinfo {volume}
  {6}},\ \bibinfo {pages} {041015} (\bibinfo {year}
  {2016}{\natexlab{d}})}\BibitemShut {NoStop}%
\bibitem [{\citenamefont {Madau}\ and\ \citenamefont
  {Dickinson}(2014)}]{Madau2014}%
  \BibitemOpen
  \bibfield  {author} {\bibinfo {author} {\bibfnamefont {P.}~\bibnamefont
  {Madau}}\ and\ \bibinfo {author} {\bibfnamefont {M.}~\bibnamefont
  {Dickinson}},\ }\href@noop {} {\bibfield  {journal} {\bibinfo  {journal}
  {Annu. Rev. Astron. Astrophys.}\ }\textbf {\bibinfo {volume} {52}},\ \bibinfo
  {pages} {415} (\bibinfo {year} {2014})}\BibitemShut {NoStop}%
\bibitem [{\citenamefont {Ivanova}\ \emph {et~al.}(2013)\citenamefont {Ivanova}
  \emph {et~al.}}]{Ivanova2012}%
  \BibitemOpen
  \bibfield  {author} {\bibinfo {author} {\bibfnamefont {N.}~\bibnamefont
  {Ivanova}} \emph {et~al.},\ }\href@noop {} {\bibfield  {journal} {\bibinfo
  {journal} {Astron. Astrophys. Rev.}\ }\textbf {\bibinfo {volume} {21}},\
  \bibinfo {pages} {59} (\bibinfo {year} {2013})}\BibitemShut {NoStop}%
\bibitem [{\citenamefont {Xu}\ and\ \citenamefont {Li}(2010)}]{Xu2010}%
  \BibitemOpen
  \bibfield  {author} {\bibinfo {author} {\bibfnamefont {X.-J.}\ \bibnamefont
  {Xu}}\ and\ \bibinfo {author} {\bibfnamefont {X.-D.}\ \bibnamefont {Li}},\
  }\href@noop {} {\bibfield  {journal} {\bibinfo  {journal} {Astrophys. J.}\
  }\textbf {\bibinfo {volume} {716}},\ \bibinfo {pages} {114} (\bibinfo {year}
  {2010})}\BibitemShut {NoStop}%
\bibitem [{\citenamefont {Inayoshi}\ \emph {et~al.}(2017)\citenamefont
  {Inayoshi}, \citenamefont {Hirai}, \citenamefont {Kinugawa},\ and\
  \citenamefont {Hotokezaka}}]{Inayoshi2017}%
  \BibitemOpen
  \bibfield  {author} {\bibinfo {author} {\bibfnamefont {K.}~\bibnamefont
  {Inayoshi}}, \bibinfo {author} {\bibfnamefont {R.}~\bibnamefont {Hirai}},
  \bibinfo {author} {\bibfnamefont {T.}~\bibnamefont {Kinugawa}}, \ and\
  \bibinfo {author} {\bibfnamefont {K.}~\bibnamefont {Hotokezaka}},\
  }\href@noop {} {\bibfield  {journal} {\bibinfo  {journal} {Mon. Not. R.
  Astron. Soc.}\ }\textbf {\bibinfo {volume} {468}},\ \bibinfo {pages} {5020}
  (\bibinfo {year} {2017})}\BibitemShut {NoStop}%
\bibitem [{\citenamefont {Visbal}\ \emph {et~al.}(2015)\citenamefont {Visbal},
  \citenamefont {Haiman},\ and\ \citenamefont {Bryan}}]{Visbal2015}%
  \BibitemOpen
  \bibfield  {author} {\bibinfo {author} {\bibfnamefont {E.}~\bibnamefont
  {Visbal}}, \bibinfo {author} {\bibfnamefont {Z.}~\bibnamefont {Haiman}}, \
  and\ \bibinfo {author} {\bibfnamefont {G.~L.}\ \bibnamefont {Bryan}},\
  }\href@noop {} {\bibfield  {journal} {\bibinfo  {journal} {Mon. Not. R.
  Astron. Soc.}\ }\textbf {\bibinfo {volume} {453}},\ \bibinfo {pages} {4456}
  (\bibinfo {year} {2015})}\BibitemShut {NoStop}%
\bibitem [{\citenamefont {Ade}\ \emph {et~al.}(2016)\citenamefont {Ade} \emph
  {et~al.}}]{Ade2015}%
  \BibitemOpen
  \bibfield  {author} {\bibinfo {author} {\bibfnamefont {P.~A.~R.}\
  \bibnamefont {Ade}} \emph {et~al.} (\bibinfo {collaboration} {Planck
  Collaboration}),\ }\href@noop {} {\bibfield  {journal} {\bibinfo  {journal}
  {Astron. Astrophys.}\ }\textbf {\bibinfo {volume} {594}},\ \bibinfo {pages}
  {A13} (\bibinfo {year} {2016})}\BibitemShut {NoStop}%
\bibitem [{\citenamefont {Hartwig}\ \emph {et~al.}(2016)\citenamefont
  {Hartwig}, \citenamefont {Volonteri}, \citenamefont {Bromm}, \citenamefont
  {Klessen}, \citenamefont {Barausse}, \citenamefont {Magg},\ and\
  \citenamefont {Stacy}}]{hartwig2016}%
  \BibitemOpen
  \bibfield  {author} {\bibinfo {author} {\bibfnamefont {T.}~\bibnamefont
  {Hartwig}}, \bibinfo {author} {\bibfnamefont {M.}~\bibnamefont {Volonteri}},
  \bibinfo {author} {\bibfnamefont {V.}~\bibnamefont {Bromm}}, \bibinfo
  {author} {\bibfnamefont {R.~S.}\ \bibnamefont {Klessen}}, \bibinfo {author}
  {\bibfnamefont {E.}~\bibnamefont {Barausse}}, \bibinfo {author}
  {\bibfnamefont {M.}~\bibnamefont {Magg}}, \ and\ \bibinfo {author}
  {\bibfnamefont {A.}~\bibnamefont {Stacy}},\ }\href@noop {} {\bibfield
  {journal} {\bibinfo  {journal} {Mon. Not. R. Astron. Soc.}\ }\textbf
  {\bibinfo {volume} {460}},\ \bibinfo {pages} {L74} (\bibinfo {year}
  {2016})}\BibitemShut {NoStop}%
\bibitem [{\citenamefont {Inayoshi}\ \emph {et~al.}(2016)\citenamefont
  {Inayoshi}, \citenamefont {Kashiyama}, \citenamefont {Visbal},\ and\
  \citenamefont {Haiman}}]{Inayoshi2016}%
  \BibitemOpen
  \bibfield  {author} {\bibinfo {author} {\bibfnamefont {K.}~\bibnamefont
  {Inayoshi}}, \bibinfo {author} {\bibfnamefont {K.}~\bibnamefont {Kashiyama}},
  \bibinfo {author} {\bibfnamefont {E.}~\bibnamefont {Visbal}}, \ and\ \bibinfo
  {author} {\bibfnamefont {Z.}~\bibnamefont {Haiman}},\ }\href@noop {}
  {\bibfield  {journal} {\bibinfo  {journal} {Mon. Not. R. Astron. Soc.}\
  }\textbf {\bibinfo {volume} {461}},\ \bibinfo {pages} {2722} (\bibinfo {year}
  {2016})}\BibitemShut {NoStop}%
\bibitem [{\citenamefont {Omukai}\ \emph {et~al.}(2005)\citenamefont {Omukai},
  \citenamefont {Tsuribe}, \citenamefont {Schneider},\ and\ \citenamefont
  {Ferrara}}]{Omukai2005}%
  \BibitemOpen
  \bibfield  {author} {\bibinfo {author} {\bibfnamefont {K.}~\bibnamefont
  {Omukai}}, \bibinfo {author} {\bibfnamefont {T.}~\bibnamefont {Tsuribe}},
  \bibinfo {author} {\bibfnamefont {R.}~\bibnamefont {Schneider}}, \ and\
  \bibinfo {author} {\bibfnamefont {A.}~\bibnamefont {Ferrara}},\ }\href@noop
  {} {\bibfield  {journal} {\bibinfo  {journal} {Astrophys. J.}\ }\textbf
  {\bibinfo {volume} {626}},\ \bibinfo {pages} {627} (\bibinfo {year}
  {2005})}\BibitemShut {NoStop}%
\bibitem [{\citenamefont {Pacucci}\ \emph {et~al.}()\citenamefont {Pacucci},
  \citenamefont {Loeb},\ and\ \citenamefont {Salvadori}}]{Pacucci2017}%
  \BibitemOpen
  \bibfield  {author} {\bibinfo {author} {\bibfnamefont {F.}~\bibnamefont
  {Pacucci}}, \bibinfo {author} {\bibfnamefont {A.}~\bibnamefont {Loeb}}, \
  and\ \bibinfo {author} {\bibfnamefont {S.}~\bibnamefont {Salvadori}},\
  }\href@noop {} {\ }\Eprint {http://arxiv.org/abs/1706.09892}
  {arXiv:1706.09892} \BibitemShut {NoStop}%
\bibitem [{\citenamefont {Inayoshi}\ \emph {et~al.}()\citenamefont {Inayoshi},
  \citenamefont {Tamanini}, \citenamefont {Caprini},\ and\ \citenamefont
  {Haiman}}]{Inayoshi2017arxiv}%
  \BibitemOpen
  \bibfield  {author} {\bibinfo {author} {\bibfnamefont {K.}~\bibnamefont
  {Inayoshi}}, \bibinfo {author} {\bibfnamefont {N.}~\bibnamefont {Tamanini}},
  \bibinfo {author} {\bibfnamefont {C.}~\bibnamefont {Caprini}}, \ and\
  \bibinfo {author} {\bibfnamefont {Z.}~\bibnamefont {Haiman}},\ }\href@noop {}
  {\ }\Eprint {http://arxiv.org/abs/1702.06529} {arXiv:1702.06529} \BibitemShut
  {NoStop}%
\bibitem [{\citenamefont {Amaro-Seoane}\ \emph {et~al.}()\citenamefont
  {Amaro-Seoane} \emph {et~al.}}]{lisa}%
  \BibitemOpen
  \bibfield  {author} {\bibinfo {author} {\bibfnamefont {P.}~\bibnamefont
  {Amaro-Seoane}} \emph {et~al.},\ }\href@noop {} {\ }\Eprint
  {http://arxiv.org/abs/1201.3621} {arXiv:1201.3621} \BibitemShut {NoStop}%
\bibitem [{\citenamefont {Nakamura}\ \emph {et~al.}(2016)\citenamefont
  {Nakamura}, \citenamefont {Ando}, \citenamefont {Kinugawa}, \citenamefont
  {Nakano}, \citenamefont {Eda}, \citenamefont {Sato}, \citenamefont {Musha},
  \citenamefont {Akutsu}, \citenamefont {Tanaka}, \citenamefont {Seto},
  \citenamefont {Kanda},\ and\ \citenamefont {Itoh}}]{decigo}%
  \BibitemOpen
  \bibfield  {author} {\bibinfo {author} {\bibfnamefont {T.}~\bibnamefont
  {Nakamura}}, \bibinfo {author} {\bibfnamefont {M.}~\bibnamefont {Ando}},
  \bibinfo {author} {\bibfnamefont {T.}~\bibnamefont {Kinugawa}}, \bibinfo
  {author} {\bibfnamefont {H.}~\bibnamefont {Nakano}}, \bibinfo {author}
  {\bibfnamefont {K.}~\bibnamefont {Eda}}, \bibinfo {author} {\bibfnamefont
  {S.}~\bibnamefont {Sato}}, \bibinfo {author} {\bibfnamefont {M.}~\bibnamefont
  {Musha}}, \bibinfo {author} {\bibfnamefont {T.}~\bibnamefont {Akutsu}},
  \bibinfo {author} {\bibfnamefont {T.}~\bibnamefont {Tanaka}}, \bibinfo
  {author} {\bibfnamefont {N.}~\bibnamefont {Seto}}, \bibinfo {author}
  {\bibfnamefont {N.}~\bibnamefont {Kanda}}, \ and\ \bibinfo {author}
  {\bibfnamefont {Y.}~\bibnamefont {Itoh}},\ }\href@noop {} {\bibfield
  {journal} {\bibinfo  {journal} {Prog. Theor. Exp. Phys.}\ }\textbf {\bibinfo
  {volume} {2016}},\ \bibinfo {pages} {093E01} (\bibinfo {year}
  {2016})}\BibitemShut {NoStop}%
\end{thebibliography}%

%\newpage

\end{document}